\documentclass[%
aps, a4paper, twocolumn,11pt
amsmath,amssymb
aps,floatfix,superscriptaddress,
pra,
nofootinbib
]
{revtex4-1}
\pdfoutput = 1
\usepackage{amsmath,amssymb}
\usepackage[toc,page]{appendix}
\usepackage[utf8]{inputenc}
\usepackage[english]{babel}
\usepackage{amsfonts}
\usepackage{qcircuit}
\usepackage{stmaryrd}
\usepackage[T1]{fontenc}
\usepackage{xkeyval}
\usepackage{tikz}
\usepackage{placeins}
\usetikzlibrary{arrows.meta}
\usepackage{graphicx}
\usepackage[export]{adjustbox}
\usepackage{soul}
\usepackage{dcolumn}
\usepackage{bm}
\usepackage{tikz}
\usetikzlibrary{calc}
\usepackage{caption}
\usepackage{subcaption}

\usepackage{amssymb}
\usepackage{overpic}
\usepackage{epstopdf}
 \usepackage[usenames,dvipsnames]{pstricks}

 \usepackage{epsfig}
 \usepackage{pst-grad} 
 \usepackage{pst-plot} 
 \usepackage[space]{grffile} 
 \usepackage{etoolbox} 
 \makeatletter 
 \patchcmd\Gread@eps{\@inputcheck#1 }{\@inputcheck"#1"\relax}{}{}
 \makeatother
\usepackage{booktabs}
\usepackage{boxhandler}
\usepackage{amsmath}
\usepackage{braket}
\usepackage{amsmath}
\usepackage{qcircuit}
\usepackage{float}
\makeatletter
\let\newfloat\newfloat@ltx
\makeatother
\usepackage{algorithm}
\usepackage{algpseudocode}
\usepackage{tikz}
\usepackage{qcircuit}

\usepackage{hyperref}
\hypersetup{
  colorlinks   = true,    
  urlcolor     = blue,   
  linkcolor    = blue,   
  citecolor    = blue     }

\usepackage{multirow}
\usepackage{enumitem}
\usepackage[normalem]{ulem}

\usepackage[amsthm]{ntheorem}

\newcommand{\F}{\mathbb{F}}

\theoremstyle{plain}
\newtheorem{theorem}{Theorem}[section]
\newtheorem{lemma}[theorem]{Lemma}
\newtheorem{corollary}[theorem]{Corollary}
\newtheorem{proposition}[theorem]{Proposition}

\theoremstyle{definition}

\newtheorem{remark}[theorem]{Remark}

\newtheorem{definition}[theorem]{Definition}

\numberwithin{equation}{section}
\newtheorem{example}[theorem]{Example}

\newcommand{\wt}{{\rm wt}}
\newcommand{\ev}{{\rm ev}}

\newcommand{\RM}{\operatorname{RM}}

\definecolor{purpleheart}{rgb}{0.41, 0.21, 0.61}

\begin{document}

\title{One Construction To Find Them: Matrix Product Quantum Codes for More Logical Qubits and Finer $Z$-Rotations}

\author{Reza Dastbasteh}
\email{rdastbasteh@unav.es}
\affiliation{Department of Basic Sciences, Tecnun - University of Navarra, 20018 San Sebastian, Spain.}
\author{Eduardo Camps Moreno}
\email{eduardo.camps-moreno@math.u-bordeaux.fr}
\affiliation{CNRS; IMB, University of Bordeaux, France.}
\author{Josu Etxezarreta Martinez}
\email{jetxezarreta@unav.es}
\affiliation{Department of Basic Sciences, Tecnun - University of Navarra, 20018 San Sebastian, Spain.}
\author{Elena Berardini}
\email{elena.berardini@math.u-bordeaux.fr}
\affiliation{CNRS; IMB, University of Bordeaux, France.}
\author{Gilles Zémor}
\email{zemor@math.u-bordeaux.fr}
\affiliation{IMB, University of Bordeaux, France. Institut universitaire de France.}

\begin{abstract}
Many protocols for preparing magic states, such as distillation and code switching, rely on quantum codes with transversal non-Clifford gates. 
Most existing constructions rely on puncturing, doubling, or numerical searches, and do not systematically provide many logical qubits together with a large distance. 
We introduce matrix product (MP) codes as a systematic framework for constructing $r$-orthogonal codes, i.e., CSS codes with a transversal logical $Z$-rotation by $\pi/2^{r-1}$ (triorthogonal for $r=3$). 
The framework combines seed codes from consecutive levels of the Clifford hierarchy into codes with many logical qubits and preserves strong orthogonality, so no Clifford corrections are needed. 
The MP structure also yields an efficient qubit recycling technique that compresses distillation protocols built from MP codes. 
Among several new high-rate codes, a $[[105,15,3]]$ strongly triorthogonal code yields, after compression to $21$ logical qubits, a factory producing $15$ $T$ states without Clifford corrections. 
We show that minimal $r$-orthogonal codes correspond to supports of Reed-Muller codewords and deduce that the shortest $r$-orthogonal codes with two logical qubits and distance three have length $2^{r+2}-4$. For every $r\ge2$ and even $k$, we construct $r$-orthogonal $[[2^r+(2^{r-1}-1)k,k,2]]$ codes that unify the H ($r=2$) and Bravyi-Haah ($r=3$) families and include the codes previously found by numerical search for $r=4$. 
They implement the logical gate without any correction operator and reduce an input error rate $p$ to $((2^{r-1}-1)k+1)p^2$ per output to leading order. 
Applying our construction to asymptotically good self-dual codes yields triorthogonal codes with overhead exponent $\gamma\to1$. 
Finally, a relaxed MP framework with an exact minimum-distance formula gives several closed-form families and explicit examples of triorthogonal codes. 
\end{abstract}

\maketitle


\section{Introduction}
Quantum error correction (QEC) is required to construct fault-tolerant quantum computers capable of running useful quantum algorithms \cite{megaquop}. 
In this context, quantum low-density parity check constructions (qLDPC) offer a viable low-overhead route towards such an objective \cite{BBcodes,xu2023constantoverheadfaulttolerantquantumcomputation,kasaicodes}. 
Importantly, the flourishing of the finite-length constructions of these codes was grounded in the hypergraph and lifted product code constructions and their variants \cite{tillich2013quantum,panteleev2022asymptotically,leverrier2022quantum}.

However, achieving universality in the fault-tolerant setting requires the preparation of high-fidelity magic states to circumvent the Eastin--Knill theorem \cite{eastinknill}. Many different approaches for preparing such resource states exist in the literature, such as magic state cultivation, distillation or code switching \cite{magicsComp,cultivation,cultivation2,distillation,heussen2025efficient,dastbasteh2026quantum}. These methods come with different space-time costs and error-suppression capabilities, suggesting that each may prove useful at some stage in the development of fault-tolerant quantum computers, or that they could be combined \cite{magicsComp}. However, reducing the overhead they incur is a fundamental necessity, especially considering that magic state preparation dominates the cost of many useful quantum algorithms \cite{femoco,litinski}. Crucially, magic state distillation (MSD) protocols reduce the logical error rates the most and, thus, are the leading candidates to prepare magic states for Teraquop fault-tolerant quantum computers. 
The trade-off comes from the huge footprint overhead that MSD protocols require to reach such fault-tolerant levels. Therefore, the design of more efficient MSD protocols is a fundamental essential in the field.

MSD protocols take $n$ copies of noisy magic states to return $k$ high-fidelity ones and are tightly related to constructing quantum error correction codes. In particular, Bravyi and Haah proposed the framework of triorthogonality to distill magic states implementing the $T$ gate \cite{bravyi2012magic}. This was later complemented with the $CS$ and $CCZ$ gates by Haah and Hastings with the notion of generalized triorthogonality \cite{haah2018codes}. 
Within this framework, assuming that Clifford operations are ideal\footnote{In practice, codes usually support ways to do fault-tolerant Clifford gates, fulfilling this notion of having ideal gates.}, an $[[n,k,d]]$ quantum code that supports a non-Clifford transversal gate is said to have a distillation overhead exponent of
\begin{equation}
    \gamma = \frac{\log{(n/k)}}{\log {d}}.
\end{equation}
The overhead exponent defines the distillation cost for an error rate below $\epsilon$ as being of the order of $\mathcal{O}(\log^\gamma(1/\epsilon))$ \cite{bravyi2012magic}. 
Recent efforts have aimed at looking for asymptotic constant overhead MSD, i.e. $\gamma\rightarrow 0$, starting from asymptotically good codes that implement the $CCZ$ gate \cite{constantCCZ,Golowich25,nguyen25} and very recently converging in proposals for the $T$ gate \cite{sanjose26,willsAsymptotic,Li26,Lin26}. 
Unfortunately, constructing good finite-length (generalized) triorthogonal codes is no trivial task, and it is of paramount importance for low-overhead fault-tolerant quantum computing. 

Most recent efforts rely on extensions of Bravyi and Haah's doubling construction \cite{bravyi2012magic,berardini2025asymptotically,jain2025transversal,dastbasteh2026quantum}, puncturing evaluation codes \cite{haah2018codes,nezami2022classification,krishna2018towards,hastings2018distillation}, or restricted numerical searches relying on symmetries, borrowed identities, noise structure, and artificial intelligence \cite{AItri,maleableTri,classificationTri,compactTri}. 
However, most of these approaches restrict to low rates or low distances due to the increasing complexity of the numerical searches and the apparent lack of structure in the codes.

In this work, we introduce matrix product (MP) codes as a constructive framework to derive $r$-orthogonal high-rate codes that can be employed for low-overhead MSD. 
While quantum MP codes have been discussed in the literature, they have been restricted to the special case of the Plotkin sum ($(u|u+v)$) construction to form valid CSS-T codes from cyclic codes \cite{QMPC}. 
However, such a restriction limits the number of logical qubits, and, as we will show in this paper, the true power of MP construction is unlocked when it is formed from larger-size matrices. 
Our framework generalizes the MP construction as a building block for $r$-orthogonal codes, combining seed codes from the $r-1$ and $r$ Clifford hierarchies to build a higher rate code belonging to level $r$. 
We formalize the MP construction for $r$-orthogonal codes, discuss the parameters and their relationship to known constructions. 
We also propose a qubit recycling technique based on \cite{recycling} that leverages the MP structure to compress MSD protocols arising from MP codes efficiently. 
Furthermore, we propose several high-rate examples useful for practical low-overhead MSD, including a $[[105,15,3]]$ strongly triorthogonal code. 
Importantly, the latter can lead to a magic state factory that requires $21$ logical qubits (after factory compression based on qubit recycling) to distill $15$ $T$-type magic states without requiring Clifford corrections. 
We also discuss how many of the codes found by means of numerical searches can be explained using our framework, showing a deep connection between good finite-length codes and the MP construction. 
Additionally, we build upon the MP construction to obtain triorthogonal codes arising from the splitting of non-triorthogonal codes into triorthogonal spaces, expanding the set of possible ingredient codes. 

Our MP construction plays a role for $r$-orthogonal codes similar to that of product constructions, such as the hypergraph product, for qLDPC codes. In particular, it combines small seed codes into larger codes with many logical qubits. 
However, whereas the hypergraph product preserves sparsity, the MP construction preserves divisibility. 
This allows the orthogonality conditions of the outer and inner codes to combine, while the number of logical qubits grows with the length of the outer code. As we will discuss, the MP framework can naturally lead to triorthogonal codes with $\gamma\rightarrow1$ asymptotically, but not below this
value, much as the hypergraph product cannot yield
asymptotically good qLDPC codes. 
All in all, our proposal sheds light on the structure of codes with the required properties for MSD, allowing systematic constructions for arbitrary lengths and distances. 
Numerical search tools can also benefit from this underlying structure to search for codes with good parameters.

\subsection{Summary of our main results}

The main focus of this paper is to introduce a systematic framework based on MP codes for constructing quantum $r$-orthogonal codes with many logical qubits, and to show that several known codes of this type, some of which were found by computer search, fit into this framework. 
Our approach is constructive and allows us to compute closed-form expressions for the parameters of the code: it yields triorthogonal codes of arbitrarily large distance, and $r$-orthogonal codes of distance two and three for every $r$, all with several logical qubits. 
Our main results are summarized below.

\begin{itemize}
\item \textbf{We identify the MP construction as a systematic framework} to combine seed quantum codes from consecutive levels of the Clifford hierarchy, boosting their orthogonality while encoding many logical qubits. 
As a particular case, a weakly $r$-orthogonal $[[n_1,1,d_0]]$ code and a weakly $(r-1)$-orthogonal $[[n_2,1,d_0]]$ code yield a weakly $r$-orthogonal $[[n_1n_2,n_1,d_0]]$ code (in the non-degenerate case), whose rate $1/n_2$ depends only on the $(r-1)$-orthogonal seed. 
The construction also preserves strong orthogonality and divisibility. 
For instance, the matrix product of the $[[15,1,3]]$ and $[[7,1,3]]$ codes is a $[[105,15,3]]$ strongly triorthogonal code. See Section \ref{sec:mpc}.

\item\textbf{We introduce a qubit recycling technique that leverages the MP structure to efficiently compress MSD protocols built from these codes.} The technique compresses the small baseline triorthogonal code and then expands it to the MP code in a form where recycling can be done directly.
This allows us to use the $[[105,15,3]]$ code to build a protocol that uses $21$ logical qubits to distill $15$ $T$-type magic states without Clifford correction. The original protocol using such code would require a footprint of $27$ logical qubits, i.e. around a $20\%$ reduction in the space footprint. See Subsection \ref{sub:qubitrecycling}.

\item \textbf{We show that minimal $r$-orthogonal codes correspond to supports of Reed--Muller codewords.} 
Using this correspondence, we show that the shortest $r$-orthogonal codes with two logical qubits and minimum distance three have length $2^{r+2}-4$, and we construct such codes for every $r$. 
This recovers the smallest triorthogonal $[[28,2,3]]$ code encoding two logical qubits, which was previously found by computer search. 
We also provide a construction for $r$-orthogonal codes of distance three and any desired number of logical qubits. See Section \ref{sec:minimalcodes}.

\item \textbf{For every $r\ge2$ and every even $k$ we construct $r$-orthogonal} $\mathbf{[[2^r+(2^{r-1}-1)k,\,k,\,2]]}$ \textbf{codes.} They unify and generalize the $[[k+4,k,2]]$ H codes ($r=2$) and the $[[3k+8,k,2]]$ Bravyi--Haah codes ($r=3$). For $r=4$ they give $[[16+7k,k,2]]$ codes with a transversal $\sqrt{T}$ gate, including the $[[30,2,2]]$, $[[44,4,2]]$ and $[[58,6,2]]$ factories previously found by numerical search in \cite{maleableTri}.
We also show that such codes realize transversal logical gates from the $r$-th level of the Clifford hierarchy and do not require any correction operator. See Appendix \ref{A:generalfamily}.

\item \textbf{We apply our MP framework to punctured and shortened Reed--Muller codes} and obtain an explicit infinite class of triorthogonal
\[
[[(2^{m_1}-1)(2^{m_2}-1),\,2^{m_1}-1,\,2^{s+1}-1]]
\]
codes for $m_1>3s$ and $m_2>2s$, with overhead exponent $\gamma=\log(2^{m_2}-1)/\log(2^{s+1}-1)$. Combined with asymptotically good self-dual codes, the same approach yields a triorthogonal family with $\gamma\to1$. See Section \ref{sec:gamma}.

\item \textbf{We relax the conditions of our MP construction} by showing that the outer matrix only needs to split into two even-triorthogonal blocks, which need not be triorthogonal together, and we determine the exact minimum distance of the resulting triorthogonal codes.
In particular, any self-orthogonal $[[n,1,d]]$ code with $n$ odd gives $[[n(d-1),2,d-1]]$ and $[[n(d+1),2,d]]$ triorthogonal codes. 
For example, $[[184,2,7]]$ is shorter than two copies of $[[95,1,7]]$, the shortest known distance-seven triorthogonal code presented in \cite{jain2025transversal}. 
Combined with the Plotkin sum construction and Reed--Muller codes, this gives the family $[[n2^{m+1},2^m,\min\{d,2^{s+2}\}]]$ for $m>3s$. See Section \ref{sec:generalMPC}.
\end{itemize}

Apart from the family with $\gamma\to1$, which relies on the existence of asymptotically good self-dual codes, 
all constructions are explicit, and their parameters follow from closed-form expressions.
Moreover, these codes benefit from the rich underlying structure of the ingredient codes and thus may offer better performance, i.e., faster decoding \cite{demarti2024decoding} and processing of logical information, than the codes found through numerical search.

\section{Background}\label{sec:background}

\subsection{Quantum codes}

A \emph{qubit} is a quantum system whose state is a unit vector in $\mathbb{C}^2$. 
We label the standard basis vectors of $\mathbb{C}^2$ by the elements of the finite field $\F_2$ and denote them by $\ket{0}$ and $\ket{1}$. 
An $n$-qubit system sits in the Hilbert space $(\mathbb{C}^2)^{\otimes n}$, and for $x\in\F_2^n$ we write $\ket{x}=\ket{x_1}\otimes\cdots\otimes\ket{x_n}$. 
These $2^n$ vectors form the so-called \emph{computational basis}. 
Quantum error-correcting codes protect quantum information by encoding $k$ \emph{logical qubits} into $n>k$ \emph{physical (data) qubits}. 
An $[[n,k,d]]$ quantum code is a $2^k$-dimensional subspace $Q\subseteq(\mathbb{C}^2)^{\otimes n}$ with minimum distance $d$, defined below. 
Such a code detects every error acting on at most $d-1$ qubits and corrects every error acting on at most $\lfloor (d-1)/2\rfloor$ qubits.

The single-qubit Pauli operators are $I$, $X$, $Z$ and $Y=iXZ$, where $X\ket{a}=\ket{a+1}$ and $Z\ket{a}=(-1)^a\ket{a}$ for $a\in\F_2$. 
The $n$-qubit Pauli group is
\[
\begin{split}
\mathcal{P}_n=\bigl\{\, & i^{c}P_1\otimes\cdots\otimes P_n \;:\; c\in\{0,1,2,3\},\ \\ & P_j\in\{I,X,Y,Z\}\,\bigr\}.
\end{split}
\]
The $n$-fold tensor products of $I,X,Y,Z$ form a basis of the complex vector space of linear operators on $(\mathbb{C}^2)^{\otimes n}$, so it suffices to correct Pauli errors \cite{degen}. The \emph{weight} of a Pauli operator is the number of qubits on which it acts nontrivially. For $x\in\F_2^n$ we write
\[
X(x)=X^{x_1}\otimes\cdots\otimes X^{x_n},\qquad Z(x)=Z^{x_1}\otimes\cdots\otimes Z^{x_n}.
\]
Both operators have weight equal to the Hamming weight $\wt(x)=\#\{i: x_i\neq0\}$, and
\[
X(x)Z(y)=(-1)^{x\cdot y}Z(y)X(x),
\]
where $x\cdot y=\sum_{i}x_iy_i\in\F_2$ denotes the \emph{Euclidean inner product} of $x$ and $y$.

A \emph{stabilizer group} is an abelian subgroup $S\subseteq\mathcal{P}_n$ with $-I\notin S$. The associated \emph{stabilizer code} is the common $+1$ eigenspace
\[
Q(S)=\bigl\{\ket{\psi}\in(\mathbb{C}^2)^{\otimes n} : M\ket{\psi}=\ket{\psi}\text{ for all } M\in S\bigr\}.
\]
If $S$ has $n-k$ independent generators, then $\dim Q(S)=2^k$ \cite{degen}. 
The minimum distance of $Q(S)$ is the smallest weight of a Pauli operator that commutes with every element of $S$ but does not belong to $\langle iI,S\rangle$; such operators are the nontrivial \emph{logical operators}.

For a binary code $C\subseteq\F_2^n$, its (Euclidean) \emph{dual} is 
\[
C^\perp=\{y\in\F_2^n : x\cdot y=0\text{ for all }x\in C\}.
\]
The Calderbank--Shor--Steane (CSS) construction~\cite{CS96,steane1996multiple} provides a fundamental bridge between classical and quantum coding theory. 
Given binary linear codes $C_2\subseteq C_1\subseteq\F_2^n$, the quantum CSS code is the stabilizer code with $X$-stabilizers $X(x)$, $x\in C_2$, and $Z$-stabilizers $Z(z)$, $z\in C_1^\perp$. These operators commute because $C_2\subseteq C_1$. 
This construction gives an $[[n,k,d]]$ quantum code with $k=\dim C_1-\dim C_2$. 
If $u_1,\dots,u_k\in C_1$ represent a basis of $C_1/C_2$, then the vectors
\[
\ket{\phi}_L=\frac{1}{\sqrt{|C_2|}}\sum_{w\in C_2}\Bigl|\,w+\sum_{j=1}^{k}\phi_ju_j\Bigr\rangle,\qquad \phi\in\F_2^k,
\]
form a basis of the code, and $X(u_j)$ acts as the logical $X$-operator on the $j$-th logical qubit. We call $u_1,\dots,u_k$ the \emph{$X$-logical operators}. The minimum distance is $d=\min\{d_x,d_z\}$, where
\[
d_x=\min_{x\in C_1\setminus C_2}\wt(x),
\qquad
d_z=\min_{x\in C_2^\perp\setminus C_1^\perp}\wt(x)
\]
are the smallest weights of $X$-type and $Z$-type logical operators, respectively. 
In MSD, the errors are of $Z$-type (see Section \ref{sec:magic factory}), so the relevant parameter there is $d_z$.

\subsection{Quantum logical gates}\label{sec:logicalGates}

Let $Q \subseteq (\mathbb{C}^2)^{\otimes n}$ be a binary quantum stabilizer code. 
We call an $n$-qubit unitary matrix $U$ a {\em logical operator} of $Q$, if $U(Q)=Q$. 
A particular case of the logical operators is the $n$-qubit unitary matrices that are tensor products of one-qubit unitary gates.
Such logical operators are inherently fault-tolerant, as a fault in a single-qubit gate cannot be propagated to other data qubits. 
A logical operator that is a tensor product of one-qubit gates will be called a {\em transversal gate}.  
There has been an increasing interest in the study of quantum codes that realize such logical gates.  

A one-qubit $Z$-rotation gate by the angle $\theta=\frac{\pi}{2^r}$, for some integer $r\ge 0$, has the form 
\[
R_Z(\theta) = e^{-i \frac{\theta}{2} Z} = \begin{bmatrix} e^{-i\theta/2} & 0 \\ 0 & e^{i\theta/2} \end{bmatrix}=e^{-i\frac{\theta}{2}}
\begin{bmatrix} 1 & 0 \\ 0 & e^{i\theta} \end{bmatrix}.
\]
For each $r\ge 0$, the gate $R_Z(\theta)$ belongs to the $(r+1)$-th level of the Clifford hierarchy. 
In particular, when $r \ge 2$, $R_Z(\theta)$ is a non-Clifford gate. 
There have been several recent developments in characterizing the conditions under which a quantum code has a transversal logical $R_Z(\theta)$ gate \cite{bravyi2012magic,webster2023transversal,leitch2025transversal}.   

We recall that the {\em Schur product} of two length $n$ binary vectors $u$ and $v$ is defined by 
\[
u \ast v=(u_1v_1,u_2v_2,\ldots,u_nv_n). 
\]
One can define the Schur product of more than two vectors analogously. For more details about the Schur product, we refer the reader to \cite{randriambololona2015products}.
This product is critical to the sufficient and necessary conditions for realizing rotation-type logical gates. 
\begin{theorem}(\cite{webster2023transversal,leitch2025transversal})\label{T:logical con}
Let $Q$ be an $[[n,k]]$ quantum CSS code of length $n$. Denote by $H_X$ the set of $X$-stabilizer generators and by $L_X=\{u_1,u_2,\ldots,u_k\}$ the $X$-logical operators. Then applying the operator
\[
\bigotimes_{i=1}^n R_Z\left(\frac{\pi}{2^{r-1}}\right)^{t_i},
\]
for some $0\le t_i\le 2^r-1$, to the data qubits realizes a logical gate \[
\bigotimes_{i=1}^k R_Z\left(\frac{\pi}{2^{r-1}}\right)^{s_i}, 
\]
for some $0\le s_i\le 2^r-1$, if and only if 
\begin{itemize}
    \item $\sum_{j=1}^{n}t_j(u_i)_j\equiv s_i \pmod{2^r}$ for each $1\le i \le k$,
    \item $\sum_{j=1}^{n}t_jv_j\equiv 0 \pmod{2^r}$ for each $v \in H_X$,
    \item the Schur product of $(t_1,t_2,\ldots,t_n)$ and any  $2\le s \le r$ number of distinct vectors in $H_X \cup L_X$ has weight divisible by $2^{r-s+1}$.
\end{itemize}
\end{theorem}
Note that the sums in the above first two bullets can be denoted by 
$\|u_i\|_t$ and $\|v\|_t$ where $t=(t_1,t_2,\ldots,t_n)$. 
This is the representation used in \cite[Definition~III.1]{haah2018towers}.

We are mostly interested in two cases. First, when $t_j=1$ for each $1\le j \le n$. 
Secondly, when the logical $X$-operators have disjoint supports.
Applying these restrictions to the result of Theorem \ref{T:logical con} reduces the conditions to the following: 
\begin{itemize}
    \item $\wt(u_i)\equiv s_i \pmod{2^r}$ for each $1\le i \le k$,
    \item $\wt(v)\equiv 0 \pmod{2^r}$ for each $v \in H_X$,
    \item the Schur product of any  $2\le s \le r$ number of distinct vectors in $H_X \cup L_X$ has weight divisible by $2^{r-s+1}$.
\end{itemize}
\begin{definition}
A quantum CSS code that satisfies the above three conditions is called 
{\em strongly $r$-orthogonal}.
\end{definition}

In the literature, the above conditions have also been employed via a small relaxation
which, usually, requires a correction operator to preserve the code space and realize the desired logical operator.  
In particular, changing the weight divisibility conditions, i.e., 
\begin{itemize}
\item in the first bullet point, only requiring $\wt(u_i)\equiv 1 \pmod 2$,
\item in the last two bullet points, only requiring divisibility by two, 
\end{itemize}
gives a weaker notion of orthogonality \cite{vuillot2022quantum,PhysRevA.92.022312,koutsioumpas2022smallest}, which we refer to as {\em weakly $r$-orthogonal}. 
As we use the definition of (weakly) triorthogonal codes ($r=3$) very frequently, we have provided the formal definition below. 

\begin{definition}(\cite{bravyi2012magic})
A binary matrix $G$ of size $m \times n$ is called triorthogonal if and only if the supports of any pair and any triple of its distinct rows have even overlap, i.e.,
$$\sum_{j=1}^nG_{a,j}G_{b,j} \equiv 0\pmod{2}$$
for all pairs of rows $1 \leq a < b \leq m$ and
$$\sum_{j=1}^nG_{a,j}G_{b,j}G_{c,j} \equiv 0\pmod{2}$$
for all triples of rows $1 \leq a < b < c\leq m.$
\end{definition}
We call the linear code generated by a triorthogonal matrix a \emph{triorthogonal subspace}. 
We can write a triorthogonal matrix as $G=\begin{bmatrix}
    G_1\\G_0
\end{bmatrix}$,
where $G_1$ consists of odd-weight rows, and $G_0$ consists of even-weight rows.
If there is no odd weight row ($G_1=\emptyset$), we call the matrix $G$ an {\em even triorthogonal} matrix.

\begin{theorem}(\cite{bravyi2012magic})\label{thm:qtoc}
Let $G=\begin{bmatrix}
    G_1\\G_0
\end{bmatrix}$ be a triorthogonal matrix, where $G_1$ consists of $k$ odd-weight rows. Let $C_1$ and $C_2$ be the linear codes generated by $G$ and $G_0$, respectively. The CSS code obtained from $C_2\subseteq C_1$ is an $[[n,k]]$ quantum triorthogonal code $Q$. Then, applying transversal $T$ to the data qubits of $Q$ realizes the transversal logical $T$ (or $T^\dagger$) on each logical qubit (up to Clifford corrections).
\end{theorem}

One can naturally extend the definition of a quantum triorthogonal code to a weakly $r$-orthogonal code.
Let $r\geq 2$ and $C\subseteq\mathbb{F}_2^n$ be a weakly $r$-orthogonal code with a basis $\beta=\{v_1,\ldots,v_k\}\subset C$. 
The matrix $A$ with rows $v_1,\ldots,v_k$ is called a weakly $r$-orthogonal matrix. 
Typically, the odd-weighted vectors of $\beta$ are written at the top of the matrix $A$. Let $C_0$ be the code spanned by the even-weighted vectors of $A$. We have that $C_0=C\cap C^\perp$ and it is uniquely determined by the code and not by the choice of the basis. Thus, the weakly $r$-orthogonal matrix $A$ defining $C$ is unique modulo $C_0$ (and reordering), i.e., if $A'$ is another weakly $r$-orthogonal matrix spanning $C$, then there is a permutation matrix $P$ such that $A-PA'$ has rows lying in $C_0$ \cite{allerton}. Finally, the corresponding quantum weakly $r$-orthogonal code arising from $C$ (or equivalently $A$) is the CSS code with $X$-stabilizers corresponding to $C_0$ and $Z$-stabilizers corresponding to $C^\perp$. The choice of the weakly $r$-orthogonal matrix $A$ affects the logical states.

\begin{remark}
    Throughout this paper, we will often drop the adjectives \emph{weakly} and \emph{strongly} for $r$-orthogonal: it is to be understood that the corresponding result applies to both cases.
\end{remark}

\subsection{Reed--Muller codes and their relatives}\label{sec: hamming and simplex}

We quickly recall some important definitions and properties of some classical codes we shall use in the paper.

We start with the definition of Reed--Muller codes. 
Let $m\geq 1$ and fix an order on $\mathbb{F}_2^m=\{P_1,\ldots,P_n\}$, $n=2^m$. Let $R_m=\mathbb{F}_2[x_1,\ldots,x_m]/\langle x_1^2+x_1,\cdots,x_m^2+x_m\rangle$ be the Boolean ring. For a polynomial $f\in R_m$, we write $\ev(f)=(f(P_1),\ldots,f(P_n))$. The Reed--Muller (RM) code of degree $s$ in $m$ variables is
    $$\RM(s,m)=\{ev(f)\ :\ \deg f\leq s\}.$$
It has length $2^m$ and minimum distance $2^{m-s}$.
Moreover, the dual of \(\operatorname{RM}(s ,m)\) is given by
\[
\operatorname{RM}(s ,m)^\perp
=
\operatorname{RM}(m-s -1,m).
\]
Let us recall some properties of RM codes that we will use later.
\begin{itemize}
    \item If $u\in R_m$ is a monomial, then $\ev(u)$ has even weight if and only if $\deg u<m$.

    \item If $v=\ev(f)$ and $w=\ev(g)$, then $v\ast w=\ev(fg)$. 

    \item For $v=\ev(f)$, we have $v_i=f(P_i)$. Thus, we can identify positions with points and thus supports of codewords with a set of points. Thus, we will refer to the support of $v$ as the set
    $$\{P\in\mathbb{F}_2^m\ :\ f(P)=1\}.$$
\end{itemize}

A well-known code which can be derived from RM codes is the Hamming code.

Let $r\ge 2$ be an integer. The binary Hamming code of order $r$ has parameters $[2^r-1,2^r-r-1,3]$ and parity-check matrix $H_r^\ast$, where $H^\ast_r$ consists of all non-zero vectors of $\F_2^r$ as its columns. 
The Hamming code of order $r$ can be obtained by puncturing $\operatorname{RM}(r-2,r)$ at one coordinate.

The dual of the Hamming code, that is, the code generated by the rows of $H^\ast_r$, is called the simplex code of order $r$. 
It has parameters $[2^r-1,r,2^{r-1}]$ and all of its non-zero codewords have weight $2^{r-1}$. 
Since puncturing and shortening are dual operations and $\operatorname{RM}(r-2,r)^\perp = \operatorname{RM}(1,r)$, the simplex code of order $r$ can be obtained by shortening $\operatorname{RM}(1,r)$ at one coordinate. 
Moreover, as it is mentioned in \cite[Lemma 2]{koutsioumpas2022smallest}, the simplex code of order $r$ is $(r-1)$-orthogonal. We will use the above information frequently throughout our discussions. 

\begin{remark}\label{R:Hadamard1}
When $r=2$, the Hamming code is the repetition code of length $3$, and the simplex code is the all-even code of length $3$. 
When $r=1$, naturally, we define the simplex code to have the generator matrix $[1]$ (the $1\times 1$ matrix with entry $1$): it is the $[1,1,1]$ code. This matches the parameters of simplex codes based on the variable $r$ above. 
\end{remark}

\subsection{Magic state distillation} \label{sec:magic factory}
Magic state distillation (MSD) protocols rely on using a set of $n$ noisy magic states to distill $k$ high-fidelity magic states by means of post-selection \cite{bravyi2012magic}. It is important to note that the MSD protocol to distill logical $\ket{T}=T\ket{+}$ states only requires checking for $Z$-errors. This occurs because, when checking the action of an $X$-error on a $\ket{T}$ type magic state, we can observe 
\begin{align*}
    X\ket{T} & = XT\ket{+} = R_Z(-\pi/2)\ket{T} \\ &= (\cos(\pi/4)I+i\sin(\pi/4)Z)\ket{T},
\end{align*}
and upon the measurement of the syndromes, this will collapse to either the correct $\ket{T}$ state or to the one experiencing phase flip errors. Therefore, it is sufficient to correct $Z$-errors. 
This is what actually happens in a practical setting since the distillation protocol operates over logical qubits; that is, the Clifford operations will be essentially ``error-free''\footnote{Recall that ``error-free'' here refers to the fact that Clifford operations are done in a fault-tolerant manner, i.e. their error rate is suppressed by means of error correction.}. The only noisy parts of the circuit will occur due to the noisy magic state injections, for which we can actually make the previous argument \cite{litinski}.

The main consequence is that, once a code that supports a transversal logical $T$ gate is built, we only care about the $X$-stabilizers to correct the phase-flip errors. However, the factory distance will be the distance of the quantum code $[[n,k,d]]$ containing $m_x$ number of independent $X$-stabilizer generators. 
This is because only the errors belonging to the normalizer minus the stabilizer set will result in logical errors, i.e., we still have stabilizer-like errors that affect the code space trivially. Now, using Litinski's framework \cite{litinski}, we can compile the magic state factory using the extended check matrix
\begin{equation*}
    H_{fact} = \begin{pmatrix}
        H_0\\H_1
    \end{pmatrix},
\end{equation*}
where $H_0$ are the $X$-stabilizers (the even triorthogonal matrix), and $H_1$ are the logical operators. 
In this way, the number of rows in this extended matrix implies how many logical qubits (or circuit width) we need for the distillation protocol,  i.e., $N=m_x+k$, and the columns indicate how many and which noisy $\pi/4$ rotations we need to perform\footnote{Note that in Litinski's framework \cite{litinski} they refer to these as noisy $\pi/8$ rotations. This comes from the fact that they do not use the factor $2$ in the $Z$-rotation definition. We stick to the usual convention we introduced in Section \ref{sec:logicalGates} for the rest of the paper.}. Thus, the number of columns determines the amount of noisy magic states consumed in the distillation process and the time overhead of the protocol. As we will discuss later, the circuit width can be reduced by means of qubit recycling techniques \cite{recycling}.

Recall that, if $\epsilon$ denotes the target output error
rate of a desired magic state, the distillation cost $C(\epsilon)$ measures the number
of noisy input magic states required per output state of
accuracy $\epsilon$.

It was shown in \cite{bravyi2012magic} that a quantum triorthogonal code with parameters $[[n,k,d]]$ has cost function
\[
C(\epsilon)
=
O\!\left(
\log^{\gamma}\!\frac{1}{\epsilon}
\right),
\]
where
\[
\gamma
=
\log_{d}\!\left(\frac{n}{k}\right).
\]
The exponent $\gamma$ characterizes the efficiency of the
distillation protocol. Smaller values of $\gamma$ correspond
to lower resource overhead and therefore to more efficient MSD protocols. 

The $[[3k+8,k,2]]$ triorthogonal family of \cite{bravyi2012magic} has the asymptotic cost value 
\[
\gamma = \log_{2} 3 \approx 1.585.
\]
The $[[15,1,3]]$ quantum Reed--Muller code has the cost of $\gamma \approx2.47$.

\section{Quantum orthogonal codes from Matrix Product codes}\label{sec:mpc}

In this section, we present a technique based on matrix product (MP) for constructing $r$-orthogonal codes. 
The application of this construction is ample, and many quantum orthogonal codes with several logical qubits can be generated this way. 
In particular, in this and subsequent sections, we will show that 
\begin{enumerate}
    \item the MP construction can be used to form several infinite classes of quantum orthogonal codes with transversal fine phase rotation (sometimes addressable) logical gates, including $T$ gates;
\item other well-known constructions such as the ``doubling construction'' and the $[[s + 4, s, 2]]$ H and $[[8 + 3s, s, 2]]$ Bravyi--Haah triorthogonal families, for even integers $s \ge 0$, can be explained using the MP approach; 
\item the latter family can be generalized to an infinite sequence of $r$-orthogonal codes with minimum distance $2$ for each $r\ge 2$;
\item the existence of an infinite family of triorthogonal codes with the resource overhead function $\gamma=\log_{d}\!\left(\frac{n}{k}\right) \to 1$.
\end{enumerate}
Our construction takes advantage of the structure of MP codes, recalled below.

\begin{definition}
Let $s$ and $\ell$ be positive integers such that $s \le \ell$.
Let $C_1, C_2, \dots, C_s \subseteq \mathbb{F}_2^m$ be $s$ binary linear codes with generator matrices $G_1,G_2,\dots,G_s$. 
Let $A = (a_{i,j})$ be an $s \times \ell$ matrix with entries in $\mathbb{F}_2$, and no zero rows; let the rows of $A$ be denoted by $R_1,\ldots,R_s$.
The matrix product (MP) code 
$$C = [C_1 \, C_2 \, \dots \, C_s] \cdot A$$ 
is defined as the 
binary linear code with generator matrix 
\begin{equation}\label{eq.mpcgen}
G = \begin{pmatrix} 
a_{1,1}G_1 & a_{1,2}G_1 & \dots & a_{1,\ell}G_1 \\
a_{2,1}G_2 & a_{2,2}G_2 & \dots & a_{2,\ell}G_2 \\
\vdots & \vdots & \ddots & \vdots \\
a_{s,1}G_s & a_{s,2}G_s & \dots & a_{s,\ell}G_s 
\end{pmatrix}=\begin{pmatrix} R_1\otimes G_1\\ \vdots \\ R_s\otimes G_s\end{pmatrix}.
\end{equation}
\end{definition}

MP codes are a generalization of well-known constructions such as tensor product codes or the $(u|u+v)$-construction \cite{mpc}. If the $s$ constituent codes $C_i$ are binary linear codes with parameters $[m, k_i, d_i]$ and the matrix $A$ has full row rank over $\mathbb{F}_2$, then the dimension $k$ of $C$ is simply 
$$k = \sum_{i=1}^s k_i.$$
The minimum Hamming distance $d(C)$ depends on the structure of $A$. Let $D_i$ be the minimum distance of the linear code in $\mathbb{F}_2^\ell$ generated by the first $i$ rows of $A$ 
$$\mathcal{R}_i = \langle R_1, R_2, \dots, R_i \rangle \subseteq \mathbb{F}_2^\ell,$$
and let $d_i = d(C_i)$ be the minimum distance of the constituent code $C_i$. 
Then, the minimum distance of the MP code satisfies
$$d(C) \ge \min_{1 \le i \le s} \{ d_i D_i \}.$$
An interesting special case is when the constituent codes are sequentially nested, i.e., $C_1 \supseteq C_2 \supseteq \dots \supseteq C_s$: the lower bound becomes tight \cite{diego}.

The simplest way to create $r$-orthogonal codes from other codes is by combining $r$-orthogonal codes themselves. From now on, if we claim that an MP code induces an $r$-orthogonal code, we also mean that the generator matrix of the form in Equation \eqref{eq.mpcgen} is an $r$-orthogonal matrix.

Before proceeding with our construction, we need the following lemma.
 \begin{lemma}
    Let $A=\begin{pmatrix} A_1\\ A_0\end{pmatrix}$ be $r$-orthogonal for some $r\ge 2$ with $A_1$ and $A_0$ being the odd-weight and the even-weight rows of the matrix, respectively. Then, there are matrices (possibly empty) $H,H'$ and $B,B'$ such that
    $$\begin{pmatrix} B'\\ H'\\ A_1\\ A_0\end{pmatrix}\ \text{ and } \begin{pmatrix} A_0\\ H\\ A_1\\ B\end{pmatrix}^t$$
are inverses of each other.
 \end{lemma}
 \begin{proof}
     Let $\begin{pmatrix} A_0\\ H\end{pmatrix}$ be a parity check matrix of the row-space of $A$. If $H$ is empty, we can ignore it; otherwise, since the hull of this code is spanned by the rows of $A_0$, then $HH^t$ is a non-singular matrix, and thus we can pick $H'=EH$ such that $H(H')^t=I$. The matrix $\begin{pmatrix} A_0\\ A_1\\ H\end{pmatrix}$ is full rank, so there exists a right inverse and, in particular, a matrix $B$ such that $A_0B^t=I$ and $A_1B^t=HB^t=0$. Finally,
     $$\begin{pmatrix} B\\ A_0\end{pmatrix}\begin{pmatrix} B^t& A_0^t\end{pmatrix}=\begin{pmatrix} BB^t&I\\ I&0\end{pmatrix},$$
then there is a matrix $D$ such that $B(B+DA_0)^t=0$. By setting $B'=B+DA_0$, we get the statement.$\hfill \square$ 
 \end{proof}  

 \begin{proposition}\label{prop:MPC}
    Let $C_1,\ldots, C_{s_1}\subseteq\mathbb{F}_2^n$ be such that they induce quantum $r'$-orthogonal codes of dimensions $k_1,\ldots,k_{s_1}$ respectively. Let $C_{s_1+1},\ldots, C_s\subseteq\mathbb{F}_2^n$ some codes and let $A\subseteq\mathbb{F}_2^{s\times N}$ be an $r$-orthogonal matrix with odd weight part of size $s_1$. Assume $r'\geq r\geq 2$. Let 
    $$D=[C_1\ \cdots\ C_s]A$$
    be the corresponding MP code and $\mathcal{R}$ the code spanned by the rows of $A$. Let $d_1=d(C_1^\perp+C_1\setminus C_1^\perp)$ and for $2\leq i\leq s_1$, let $d_i=d(C_i^\perp+C_i)$. Take $\mathcal{R}_i$ as the span of the first $i$ rows of $A$. Then, $D$ induces an at least $r$-orthogonal quantum code of length $nN$, dimension $k_1+\cdots+k_{s_1}$ and distance at least the minimum between
    $$\min_{1\leq j\leq s_1}\{d((\mathcal{R}^\perp+\mathcal{R}_j)\setminus \mathcal{R}^\perp)d_j\}$$ and 
    $$\min_{1\leq j\leq s-s_1}\{d((\mathcal{R}^\perp+\mathcal{R}_{s_1}+\mathcal{B}_i)\setminus(\mathcal{R}^\perp+\mathcal{B}_{i-1}))d(C_{s_1+j}^\perp)\}.$$
   
   \noindent where each $\mathcal{B}_{i}$ is defined following the convention of $\mathcal{R}_{i}$.
 \end{proposition}

\begin{proof}

    Let $A=\begin{pmatrix} A_1\\ A_0\end{pmatrix}$, where $A_1$ has rows of odd weight and $A_0$ has rows of even weight.

    Using the matrices from the previous lemma, $D$ can also be seen as the code
    $$[0\ \cdots\ 0\ C_1\ \cdots\ C_s]\begin{pmatrix}  B'\\ H'\\ A_1\\ A_0\end{pmatrix}$$
and thus its dual can be regarded as
    $$[\mathbb{F}_2^n\ \cdots\ \mathbb{F}_2^n\ C_1^\perp\ \cdots\ C_s^\perp]\begin{pmatrix} A_0\\ H\\A_1\\B\end{pmatrix}.$$
    
    The bound for the minimum distance follows the same proof as the original one for the MP construction, but considering that we need to compute the distance $d(D^\perp+D\setminus D^\perp)$. We replicate the argument below for the sake of completeness. 
    Let $$z=[v_1\ \cdots\ v_N] \begin{pmatrix} A_0\\ H\\A_1\\B\end{pmatrix}$$ be a codeword in $(D^\perp+D)\setminus D^\perp$. This implies that there is $1\leq j\leq s_1$ such that $v_{N-s+j}\in (C_j^\perp+C_j)\setminus C_j^\perp$ and thus it cannot be $0$. Let $1\leq i\leq N$ such that $v_h=0$ for any $h>i$; note that $i\geq N-s+1$. Thus, for each $h\in\mathrm{supp}(v_i)$, the $h$-column of $z$, $z_h$, is an element of the code spanned by the first $i$-rows of $\begin{pmatrix} A_0\\ H\\A_1\\B\end{pmatrix}$, thus, if $i\leq N-s+s_1$, then we get $$wt(z)\geq wt(v_i)d((\mathcal{R}^\perp+\mathcal{R}_j)\setminus\mathcal{R^\perp}).$$ Otherwise, we get $$wt(z)\geq wt(v_i)d((\mathcal{R}^\perp+\mathcal{R}_{s_1}+\mathcal{B}_i)\setminus(\mathcal{R}^\perp+\mathcal{B}_{i-1})),$$
from which we get the conclusion.$\hfill \square$
\end{proof}

\begin{remark}
 Observe that the bound does not consider the distance of $C_i+C_i^\perp$ for those codes multiplied by rows of $A_0$; that would be the usual bound for the corresponding $r$-orthogonal code. Thus, even if the constituent codes have trivial distance, we may recover a code with a better minimum distance.
\end{remark}

\begin{remark}
    The previous bound on the minimum distance requires the choice of a matrix $B$ that in turn depends on the choice of a parity check matrix for the code spanned by $A$. While the resulting $r$-orthogonal code is independent of such choices, the bound above can be improved by choosing the matrix $B$ that maximizes the respective distances.
\end{remark}

\begin{remark}
    Similarly to the classical MP case, the bound for the minimum distance is a rough approximation that depends just on the weight of the last non-zero column appearing in the matrix to build a codeword. A finer analysis would require further knowledge of the minimums arising from supports of the constituent codes and the relative distances for the subcodes spanned by the matrix. Later, we will restrict ourselves to the case of $r$-orthogonal matrices with a single row of odd weight, which allows us to provide better bounds and, sometimes, explicit distances.
\end{remark}

\begin{remark}
    Naturally, one would like to pick a non-singular $r$-orthogonal matrix $A$, but such a matrix can only be the identity \cite{allerton}.
\end{remark}
A more gentle bound for the minimum distance is the following.

\begin{corollary}\label{cor:simpler bound}
   Let $C_1,\ldots, C_s\subseteq\mathbb{F}_2^n$ such that they induce quantum $r'$-orthogonal codes of dimensions $k_1,\ldots,k_s$ and let $A\subseteq\mathbb{F}_2^{s\times N}$ be an $r$-orthogonal matrix with odd weight part of size $s_1$. Assume $r'\geq r$. Let 
    $$D=[C_1\ \cdots\ C_s]A$$
be the corresponding MP code and $\mathcal{R}$ the code span by the rows of $A$. Let $\delta=\min\{d(C_i+C_i^\perp)\ :\ 1\leq i\leq s_1\}$. Then the distance of the $r$-orthogonal code arising from $D$ is lower bounded by
    $$\min\{\delta d((\mathcal{R}+\mathcal{R}^\perp)\setminus \mathcal{R}^\perp),d(C_i^\perp)\ :\ s_1+1\leq i\leq s\}.$$
\end{corollary}

\begin{example}\label{ex:2dimtri}
    With the previous proposition, we can create interesting triorthogonal codes from trivial ones.

    Let $C_1=\langle (1,0,0,0),(0,0,0,1),(0,1,1,0)\rangle$ and $C_2=\langle (0,0,1,1),(1,1,1,1)\rangle$. Let
    $$A=\begin{pmatrix} 1\\ H_3^\ast\end{pmatrix}.$$

    Observe that $C_1$ induces a distance $1$ triorthogonal code, while $C_2$ induces a $0$-dimensional triorthogonal code. However, the code
    $$[C_1\ C_2\ C_2\ C_2]A$$
induces a $2$-dimensional code that, by the previous proposition, has distance at least $2$ and $\geq 2$-orthogonality. We prove later that, in fact, this code is $3$-orthogonal of distance $3$ and that it is the shortest code of dimension $2$ with these properties.
\end{example}

\begin{remark}
 One would be tempted to split the $r$-orthogonal codes $C_1,\ldots,C_s$ into just the even and odd parts, with respect to how they are going to multiply the rows of $A$; this makes the codes $C_1,\ldots,C_{s_1}$ linear complementary dual. However, if that is the case, then the bound in Proposition \ref{prop:MPC} coincides with the bound in Corollary \ref{cor:simpler bound} and it is an equality, i.e., the minimum distance of the code is
 $$\min\{d(B+B^\perp),d(C_i^\perp)\ :\ s_1+1\leq i\leq s\}.$$ 

 Thus, if we let $C_1'=\langle (1,0,0,0),(0,0,0,1)\rangle$ and we take $C_2$ and $A$ as in Example \ref{ex:2dimtri}, we get that the code $[C_1'\ C_2]A$ induces a triorthogonal code with distance exactly 2. 
\end{remark}

A simple way to increase the length while preserving the dimension and the distance is the following.

\begin{corollary}\label{cor:incr}
    Let $A=\begin{pmatrix} \mathbf{1}\\ H\end{pmatrix}\subseteq\mathbb{F}_2^{s\times N}$ be an $r$-orthogonal matrix, with $s\geq r+2$ odd. Let $C_1\subseteq\mathbb{F}_2^n$ be an $(r+1)$-orthogonal code of dimension $k_1$ and $C_2,\ldots,C_{s}\subseteq\mathbb{F}_2^n$ be codes such that $C_{i_1}\ast \cdots \ast C_{i_{r+1}}$ is even for any $2\leq i_1<\cdots<i_{r+1}\leq s$. Then
    $$D=[C_1\ \cdots\ C_s]A$$
is an $(r+1)$-orthogonal code of dimension $k_1$. If, furthermore, $d(C_i^\perp)\geq d(C_1+C_1^\perp)$ for any $2\leq i$, then $d((D^\perp+D)\setminus D^\perp)\geq d((C_1+C_1^\perp)\setminus C_1)$.

\end{corollary}

\begin{proof}
    The code is at least $r$-orthogonal by Proposition \ref{prop:MPC}. 
    We choose arbitrary $r+1$ rows of the generator matrix in the form of Equation \eqref{eq.mpcgen} of $D$, this is $v_{i_1}\otimes R_{i_1}\ldots,v_{i_{r+1}}\otimes R_{i_{r+1}}$, where $i_1,\ldots,i_{r+1}\leq s$ and each $v_{i_j}$ is a row of $G_{i_j}$, this is a generator matrix of the code $C_{i_j}$.
    If there is at least a pair $(j,j')$ such that $i_j=i_{j'}$, then by the $r$-orthogonality of $A$, we know that $(v_{i_1}\ast\cdots\ast v_{i_{r+1}})\otimes (R_{i_1}\ast\cdots\ast R_{i_{r+1}})$ is even because the second factor of the tensor product is even if at least one of the $i's$ is not $1$; otherwise, the $(r+1)$-orthogonality of $C_1$ guarantees that the first factor is even.

    So, let us assume that all the $i$'s are different. 
    If $i_j=1$ for some $j$, then $R_{i_j}=\mathbf{1}$ and the $r$-orthogonality of $A$ implies that the same factor is still even. Finally, if none of the $i$'s is 1, then $v_1\ast\cdots\ast v_{r+1}\in C_{i_1}\ast\cdots C_{i_{r+1}}$ and thus it is an even codeword. This finishes the proof of $(r+1)$-orthogonality. 
    The conclusion about the minimum distance follows from Corollary \ref{cor:simpler bound}.\hfill $\square$
\end{proof}

As mentioned, a tighter bound for the minimum distance will require further knowledge of the supports of the matrices with columns in $C_1^\perp,\ldots,C_s^\perp$. An example where we can compute the minimum distance explicitly is the following.

\begin{theorem}\label{T:newmatrix product}
    Let $A=\begin{pmatrix}\mathbf{1}\\ H\end{pmatrix}\in\mathbb{F}_2^{s\times N}$, $N$ odd, be an $r$-orthogonal matrix. Let $C_1\subseteq\mathbb{F}_2^n$ be an $r$-orthogonal code and $C_2\subseteq\mathbb{F}_2^n$ be a $2$-orthogonal code. Let $\mathcal{R}$ be the code spanned by the rows of $A$ and write
$$\delta=d((\mathcal{R}^\perp+\mathcal{R})\setminus\mathcal{R}^\perp)$$
and 
    $$\delta'=d(((C_1^\perp+C_1)\cap C_2^\perp)\setminus C_1^\perp).$$

    If $d((C_1^\perp+C_1)\setminus C_1^\perp)=1$, then there is an $r$-orthogonal quantum code with parameters $$[[nN, \dim(C_1/(C_1\cap C_1^\perp), \min\{\delta,\delta'\}]].$$
\end{theorem}

\begin{proof}
    Let $D$ be the code $[C_1\ C_2\ \cdots\ C_2]A$. Proposition \ref{prop:MPC} proves the claims about the length, the distance and the $r$-orthogonality of $D$. Let us prove the minimum distance.

    Consider $z=(z_1,\ldots,z_N)\in\mathbb{F}_2^{nN}$ be a codeword in $(D^\perp+D)\setminus D^\perp$, or, equivalently, $z=\sum_{i=1}^N e_i\otimes z_i$. Let $t_i=\sum_{j=1}^N a_{ij}z_j^t$, where $A_{ij}$ is the $(i,j)$-th entry of $A$. Let $G$ be a generator matrix of $C_2$. Given that $(R_i\otimes G)z^t=\sum_{j=1}^N a_{ij}(Gz_j^t)=G(\sum_{j=1}^N a_{ij} z_j^t)$ and $z\in D^\perp$, then we have $t_i\in C_2^\perp$ for any $2\leq i\leq N$; and a similar reasoning proves that $t_1\in C_1^\perp+C_1$; furthermore, if $z\notin D^\perp$, then $t_1\notin C_1^\perp$. 
    Now, if $wt(z)>\delta$ then we are done, so let us assume $wt(z)<\delta$. If this is the case, the cardinality of $J=\{j\in[N]\ :\ z_j\neq 0\}$ is smaller than $\delta$ and the shortening of $\mathcal{R}^\perp$ outside those positions has to be composed of even-weight codewords (otherwise, their weight would be bigger than $\delta$, a contradiction), which proves that the vector $\mathbf{1}_J$ (the vector with zeros everywhere but in the positions of $J$) is in the puncturing of $H$ in the same positions. That is, there are $\lambda_2,\ldots,\lambda_{N}\in\mathbb{F}_2$ such that $\sum_{i=2}^{N} \lambda_i a_{ij}=1$ for each $j\in J$. 
    This proves that
    $$t_1=\sum_{j=1}^N z_j^t=\sum_{i=2}^N \lambda_i t_i^\perp\in C_2^\perp.$$
    Thus $wt(z)=\sum_{i=1}^N wt(z_i)\geq wt(t_1)\geq wt((C_1^\perp+C_1)\cap C_2^\perp\setminus C_1^\perp)$, from where we get the conclusion.
$\hfill \square$    
\end{proof}

Next, we present two special cases of the above result which allow us to build orthogonal codes in simple steps. 

\begin{corollary}\label{C:newMPC1}
    Let $Q_1$ and $Q_2$ be weakly $r$-orthogonal and $(r-1)$-orthogonal quantum codes with parameters $[[n_1,1,d_1]]$ and $[[n_2,1,d_2]]$, where $n_2$ is odd. Then there exists a quantum weakly $r$-orthogonal code $Q$ with parameters 
    $$[[n_1n_2,n_1,\min\{d_2,d(C)\}]]$$
where $C$ is the binary code corresponding to the $Z$-stabilizer and logical operators of $Q_1$. 

Furthermore, if $Q_1$ and $Q_2$ are strongly $r$- and $(r-1)$-orthogonal, respectively, then $Q$ is also strongly $r$-orthogonal.
\end{corollary}

\begin{proof}
    Let $C_2$ be the code corresponding to the $X$-stabilizers of $Q_1$ and let $H$ to be a generator matrix of $Q_2$, respectively, and apply the previous theorem to $A=\begin{pmatrix} \mathbf{1}\\ H\end{pmatrix}$ and the codes $\mathbb{F}_2^n$ and $C_2$. \hfill$\square$
\end{proof}

Applying the above result to the $[[7,1,3]]$ and $[[15,1,3]]$ quantum Hamming and Reed--Muller codes leads to a $[[105,15,3]]$ triorthogonal code.
This code has a better rate than the $[[114,14,3]]$ triorthogonal code considered in \cite{haah2018codes} obtained from puncturing a Reed--Muller code. 
Some other distance $3$ triorthogonal codes obtained via MP are presented in Table \ref{tab:MPC-RM}.

The distillation cost of the $[[105,15,3]]$ triorthogonal code is  
\[
\gamma = \log_{3} 7 \approx 1.77.
\]
Going back to the factory construction discussion given in Section \ref{sec:magic factory}, we can compare the specific factories constructed with the MP construction against some other typical benchmarks. 
In particular, we get $[[105,15,3]]$ with $12$ $X$-checks. Hence, we would use $N=27$ logical qubits for $105$ noisy $\pi/4$ rotations.

The typical $[[15,1,3]]$ Reed--Muller code, i.e. the smallest code with a transversal $T$ gate and distance $3$, requires $5$ logical qubits to distill a single magic state with distance $3$ using $15$ rotations. Comparing them, we see that our factory has a higher space efficiency, being able to get $15$ magic states using $N=27$ logical qubits, i.e. a rate of almost $1/2$, while the standard factory gets a single one using $5$, i.e., a rate of $0.2$. 
In the next section, we apply the recycling technique to further optimize the distillation footprint by reducing the number of logical qubits from $N=27$ to $N=21$. 
The comparison is not even equivalent to parallelizing $15$ $[[15,1,3]]$ factories. 
In the latter case, we need $75$ logical qubits ($60$ with qubit recycling as each can be done using $4$), but much less time overhead, i.e. $15$ rotations compared to $105$.

Next, let $A$ be the generator matrix of the Reed--Muller code $\operatorname{RM}(1 ,4)$ and $Q$ be the quantum self-orthogonal $[[17,1,5]]$-code. 
Then, Corollary \ref{C:newMPC1} implies the existence of a triorthogonal code with parameters $[[272,16,4]]$. 

Moreover, the smallest currently known quantum triorthogonal (also triply-even) and 2-orthogonal (also doubly-even) codes with minimum distance $5$ have parameters $[[49,1,5]]$ and $[[17,1,5]]$. 
Both codes are degenerate, and they have classical distance of $4$. 
Hence, applying Theorem \ref{T:newmatrix product} implies the existence of a quantum triorthogonal (also triply even) code with parameters $[[833,49,4]]$.

\begin{table}[t]
\centering
\caption{Parameters of $d=3$ triorthogonal codes obtained from MP of triorthogonal $\operatorname{RM}(1,m)$ (or punctured)  with $m\ge 4$ codes with Steane code. Here $N$ shows the circuit width of the magic factory, where $N=k+\dim(X\text{-stabilizer})$.}
\label{tab:MPC-RM}
\begin{tabular}{lcr}
$n$ & $k$ & $N$  \\
\toprule
$7\times(2^m-1)$& \ $(2^m-1)$ \ & $(2^m-1)+3m$ \\
$7\times2^m$&$2^m$& $2^m+3(1+m)$\\
\hline
&Examples&\\
\hline
105 & 15 & 27 \\
112 & 16 & 31 \\
217 & 31 & 46 \\
224 & 32 & 50 \\
441 & 63 & 81 \\
448 & 64 & 85 \\
889 & 127 & 148 \\
896 & 128 & 152 \\
\bottomrule
\end{tabular}
\end{table}

A generalization of the result of Theorem \ref{T:newmatrix product} restricted to the triorthogonal codes is presented below. 
Later we show applications of this result in building infinite families of quantum triorthogonal codes. 
We represent the set of indices of the zero columns of a matrix $B$ by $B^0$. 
We also represent the vector space generated by $e_j$ (the standard basis) for each $j \in B^0$ by $\overline{B^0}$.

\begin{proposition}\label{P:twotriorthogonal}
Let $C$ be an odd length $n$ self-orthogonal code, and let $Q$ be the $[[n,1,d>1]]$ quantum CSS code corresponding to $C\subseteq C \oplus \{0_n,1_n\}$.
Let $m$ be a positive integer and $A$ and $B$ be two even binary triorthogonal matrices with $m$ columns. 
Then, there exists a quantum triorthogonal code with parameters
\[
[[nm,|B^0|, \min \{d,d(D^\bot \setminus D'^\bot)\}]],
\]
where $D$ is the binary linear code generated by 
$\begin{bmatrix}
A\\B    
\end{bmatrix}$ and $D'=D+\overline{B^0}$.
\end{proposition}

\begin{proof}
    Let $G$ be a generator matrix of $C$, $C_1$ be the row-span of $B$ plus $\overline{B^0}$, and $C_2$ be the row-span of $A$. Applying the previous theorem to the matrix $\begin{pmatrix} 1\\ G\end{pmatrix}$ and the codes $C_1$ and $C_2$ gives the desired quantum code.\hfill$\square$
\end{proof}

\subsection{Distillation footprint compression by means of qubit recycling 
}
\label{sub:qubitrecycling}
As described before, triorthogonal codes can be used to define magic state distillation factories with a spatial footprint (or circuit width) equal to the number of $X$-stabilizers plus the number of encoded logical qubits, i.e. $N=\dim(X\text{-stabilizer})+k$ \cite{litinski}. However, the spatial footprint can be reduced by means of so-called \textit{qubit recycling} techniques \cite{recycling}, based on reusing logical qubits, noticing that some of them can be either initialized later or measured in advance. 
This protocol compression is achieved by row operations and column permutations on the triorthogonal matrix defining the factory and, while useful to save logical qubits, it is an NP-hard problem. 
Therefore, finding suitable qubit recycling protocols for large triorthogonal matrices is a hard search task \cite{recycling}.

\begin{figure*}[htbp]
    \includegraphics[width=\textwidth]{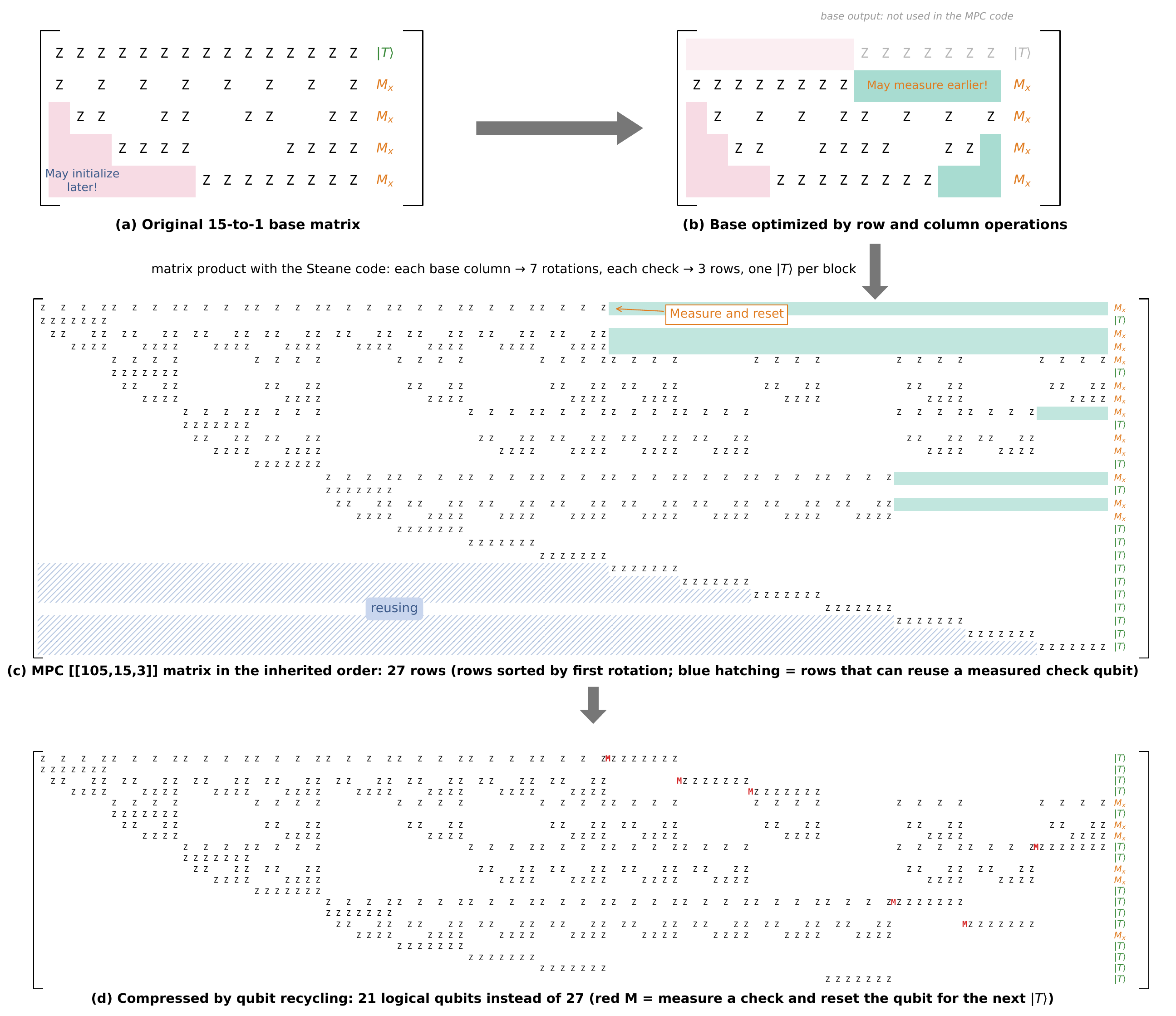}
    \caption{\textbf{Qubit recycling for a factory using MP structure.} The columns refer to $\pi/4$ noisy rotations over the logical qubits given by the rows. a) The baseline $[[15,1,3]]$ Reed--Muller code. b) Qubit-recycled version of the Reed--Muller code. c) Matrix product with the Steane code. d) Recycled factory for the $[[105,15,3]]$ from the recycled version of the base triorthogonal code. This figure is motivated by Figure 7 in \cite{recycling}.}
    \label{fig:105recycle}
\end{figure*}

Our proposed MP construction for triorthogonal codes comes from the matrix product of a triorthogonal code with a self-orthogonal code. 
This provides a clear structural insight into the block structure of the resulting high-rate triorthogonal code and, thus, presents an opportunity to systematically propose qubit recycling techniques easier to handle than raw search over the resulting matrix. 
For doing so, we leverage the fact that zero-valued sub-blocks of the MP triorthogonal matrix can be analyzed from the baseline triorthogonal matrix used for the construction. 
This allows us to find a qubit recycling method for such a building-block code that is small, and then perform the required matrix product to obtain a compressed factory for the bigger MP construction. 
While this disregards further potential general compressions coming from the structure of the self-orthogonal inner code, i.e., it is not optimal in general, it makes the problem tractable. 

The factory compression technique making use of this systematic structure of triorthogonal codes built from the MP construction can be summarized in the following steps:
\begin{enumerate}
    \item \textbf{Recycle the base: }find a qubit recycling schedule by means of row operations and column permutations for the base triorthogonal code.  
    The output row will not be used. 
    Note that this can be done with the techniques explained in \cite{recycling}.
    \item \textbf{Matrix product: }build the MP code by integrating the self-orthogonal code. 
    This is equivalent to adding the baseline self-orthogonal code's $X$-stabilizers in the non-zero positions of the compressed schedule of the triorthogonal code.
    \item \textbf{Add the logical outputs: }output $t$ is the self-orthogonal code's logical on block $t$.
    It starts at block $t$ and stays alive until the end, regardless of where recycling happens.
    \item \textbf{Assign the qubits: }each output that starts at block $t$ takes a qubit freed by a check that has already been measured, if one is free. 
    Otherwise, a new qubit is assigned for it.
\end{enumerate}

The resulting factory footprint follows directly from the recycled base. 
Let the base triorthogonal code have $m_{\mathrm{out}}$ $X$-stabilizer checks on $n_{\mathrm{out}}$ columns (code length), and let check $i$ in the recycled schedule have its first and last non-zero entries in columns $f_i$ and $\ell_i$, respectively, so that the number of base checks alive at column $t$ is $s(t)=|\{i: f_i\le t\le \ell_i\}|$. Let the self-orthogonal code have $m_{\mathrm{in}}$ $X$-stabilizers. Each base column becomes a block, each base check becomes $m_{\mathrm{in}}$ rows alive on the same blocks, and at block $t$ the outputs of blocks $1,\dots,t$ are alive. 
Hence
\begin{equation*}
N \;\le\; \max_{1\le t\le n_{\mathrm{out}}}\Bigl[\,m_{\mathrm{in}}\,s(t)+t\,\Bigr],
\end{equation*}
compared with $N=n_{\mathrm{out}}+m_{\mathrm{in}}m_{\mathrm{out}}$ without recycling. Note that, for the MP code, the number of logical qubits is the length of the triorthogonal code, $k=n_{\mathrm{out}}$, while the number of checks is the product of the numbers of checks of both constituent codes, $m_\mathrm{MPC}=\dim(X\text{-stabilizer}) = m_{\mathrm{in}}m_{\mathrm{out}}$.
The bound counts each row as alive during whole blocks, so ordering the rotations inside each block can only lower it. Every base check closed early saves $r_{\mathrm{in}}$ logical qubits in the larger factory. 
For a fixed column order, the best choice of row operations gives $s(t)=\operatorname{rank}H_{\ge t}+\operatorname{rank}H_{\le t}-m_\mathrm{out}$, where $H$ is the base check matrix restricted to the columns from $t$ onward or up to $t$. 

Applying this technique, we can provide a practical example using one of the codes we proposed in the previous section. 
Recall that, thanks to Corollary \ref{C:newMPC1}, we are able to construct the $[[105,15,3]]$ triorthogonal code by means of the matrix product of the $[[15,1,3]]$ Reed--Muller code and the Steane code. Importantly, the resulting code is strongly triorthogonal, inheriting such a property from the Reed--Muller seed, indicating that no Clifford correction is needed for the factory, which requires $N=27$ logical qubits as a physical footprint. We showcase the qubit recycling process that we proposed before in Figure \ref{fig:105recycle}. 
Here, the columns mean the noisy $\pi/4$ rotations required by the factory over the logical qubits given by each row. The process starts from the baseline $[[15,1,3]]$ Reed--Muller code that requires $5$ logical qubits, shown in panel (a). 
After qubit recycling by means of row operations and column permutations, the factory can be compressed to $4$ logical qubits\footnote{It would measure the first check and reuse that logical qubit as the output after step $8$}, shown in panel (b). 
The next step, shown in panel (c), relies on using the MP of the compressed factory with the Steane code, discarding the logical output in panel (b) and assigning the $15$ logical qubits introduced. Finally, the outputs that begin after the checks measured before the end of the protocol can be assigned to those, recycling the already measured qubits. 
Note how those positions correspond to the same ones as the original compressed Reed--Muller code. The final compressed distillation protocol can be seen in panel (d) of Figure \ref{fig:105recycle}, where the red $M$ symbols indicate that the checks are measured at such a point before being reinitialized. 

Overall, the obtained protocol requires $N=21$ logical qubits to distill $15$ magic states without requiring any Clifford correction. Furthermore, following the approach of \cite{bravyi2012magic}, there are $350$ weight-three logical $Z$-operators (undetectable errors). Hence, by post-selecting non-trivial syndromes, the suppression of the probability of ``at least one'' wrong output magic state is given by $350p^3+\mathcal{O}(p^4)$. Therefore, we can achieve a very efficient MSD protocol using our MP construction.

To sum up, triorthogonal codes built from our MP framework provide benefits for applying qubit recycling techniques in an efficient manner, allowing us to relax the optimization problem to the baseline triorthogonal matrix. 
While the recycling may not be optimal in general when following this process, it makes it easily tractable, and we have already observed that it can lead to substantial space reductions. 
Furthermore, the inherited properties of strong triorthogonality allow us to build factories that do not require Clifford corrections, indicating that the time overhead can also be more efficient.

\section{Shortest orthogonal codes of distance $2$ and $3$}\label{sec:minimalcodes}

In this section, we study $r$-orthogonal codes of distances $2$ and $3$ that have the shortest length for dimension $2$, and study their relation with Reed--Muller codewords. 

First, such codes have to be non-degenerate.

\begin{proposition}
    Let $n$ be the minimum length of a quantum $r$-orthogonal code of dimension $k$ and distance $d$. If $C_2\subseteq C_1$ induce an $r$-orthogonal code $Q$ with parameters $[[n,k,d\geq 2]]$, then $d(C_1^\perp)\geq 3$.
\end{proposition}

\begin{proof}
    Let $C_2\subseteq C_1$ be the code defining $Q$ and assume that $d(C_1^\perp)\leq 2$. If $d(C_1^\perp)=1$, then both $C_2$ and $C_1$ are degenerate, and after puncturing in the position of degeneracy, we get that the punctured code $Q'$ is $r$-orthogonal of the same distance as $Q$, contradicting its minimality. Thus, $d(C_1^\perp)=2$. But this implies that both codes $C_2\subseteq C_1$ have repeated columns in the positions of $\mathrm{supp}(v)$. Since $C_2$ cannot be degenerated, we can pick an $r$-orthogonal basis for $C_1$ such that all the odd-weighted vectors have zeros in the support of $v$. Furthermore, if $C'$ is the shortening of $C_1$ in the non-zero positions of $v$, we have that $C'$ induces an $r$-orthogonal code of dimension $\dim Q$, length $n-2$ and at least the same distance as $Q$, which is a contradiction. \hfill$\square$
\end{proof}

If $C_1$ induces a minimal $r$-orthogonal code, then we prove below that a generator matrix of $C_1$ has columns that can be seen as points supporting a codeword of a Reed--Muller code.

\begin{proposition}
    Let $G=\begin{pmatrix} G_1\\ G_0\end{pmatrix}$ be a matrix of full rank $k=k_1+k_0$, with $\mathrm{rank}\ G_1=k_1$, the rows of $G_1$ all of odd weight and the rows of $G_0$ all even. Assume that $G$ has no repeated columns. The following are equivalent:
    \begin{enumerate}
        \item $G$ is an $r$-orthogonal matrix.
        \item The matrix
        $$G_e=\begin{pmatrix}
        I_{k_1}&G_1\\
        0&G_0\end{pmatrix}$$
is $r$-orthogonal, and all its rows are even.
        \item Let $\mathcal{P}\subseteq\mathbb{F}_2^{k}$ be the columns of $G_e$. Then there is a polynomial $f\in R_k$ of degree $\leq k-r-1$ such that the set of zeros of $f$ is $\mathbb{F}_2^k\setminus \mathcal{P}$. 
    \end{enumerate}
\end{proposition}

    \begin{proof}
        The first and the second item are trivially equivalent. For the third, observe that for any $g\in R_k$ of degree $r$, we have
        $$G_e\text{ is }r\text{-orthogonal}\Longleftrightarrow \sum_{P\in \mathcal{P}}g(P)=0.$$

        But the last is equivalent to saying that $\mathbf{1}_\mathcal{P}\in (\RM(r,k))^\perp=\RM(k-r-1,k)$, where
        $$(\mathbf{1}_\mathcal{P})_i=\begin{cases} 1& P_i\in \mathcal{P}\\ 0&\text{otherwise}\end{cases}.$$ As such, there is a polynomial $f$ of degree up to $k-r-1$ such that $\mathcal{P}$ is the support of $\ev(f)$.\hfill$\square$
    \end{proof}

\begin{remark}
    The previous result is the generalization of Lemma 5 in \cite{nezami2022classification}.
\end{remark}

We may assume that $k\geq r+2$ to have at least an affine space of dimension $r+1$ that does not include the $0$ vector (otherwise, the code would be degenerate). For the binary Reed--Muller codes $RM(d,m)$, we know that:
\begin{itemize}
    \item They are generated by their minimal weight codewords. 

    \item Any minimal weight generated codeword is supported on a $(m-d)$-dimensional affine space.

    \item Any codeword of $\RM(d,m)$ of weight $<2^{m-d+1}$ is the sum of two minimum weight codewords \cite{kasami}. 
\end{itemize}

Any RM codeword is the symmetric difference of such affine spaces. Furthermore, since we would like to have full-rank matrices, we can assume that 
$$e_1,\ldots,e_k\in \mathcal{P}\subseteq\mathbb{F}_2^k.$$

The second item above gives the known fact that a minimal $r$-orthogonal code of dimension $1$ and distance at least $3$ has length $2^{r+1}-1$, and it is the realization of the space $e_1+\langle e_2,\ldots,e_{r+2}\rangle$, where $e_i\in\mathbb{F}_2^{r+2}$ is the point with all zero entries except in position $i$.

The previous observations allow us to prove that any $r$-orthogonal code of dimension $2$ and distance $3$ has length either $\geq 2^{r+2}$ or  $2^{r+2}-4$. 

\begin{theorem}
Let $Q$ be an $r$-orthogonal code of parameters $[[n,2,3]]$. Then $n\geq 2^{r+2}-2$ or $n=2^{r+2}-4$. 
\end{theorem}

\begin{proof}
    Let $C_0\subseteq C_1$ such that $Q(C_1,C_0)$ is $r$-orthogonal of minimal length. Let $G$ be an $r$-orthogonal generator matrix of $C_1$ and let $\mathcal{P}$ be the columns of $G_e$ obtained from $G$ as before, where $k$ is the dimension of $C_1$. We get that $\mathcal{P}$ is the support of a codeword of $\RM(k-r-1,k)$ and thus $|\mathcal{P}|\geq 2^{r+1}$. Now, if $|\mathcal{P}|\geq 2^{r+2}$, we are done. If $|\mathcal{P}|<2^{r+2}$, by \cite{kasami}, we have that $\mathcal{P}=\mathcal{P}_1\Delta \mathcal{P}_2$ and $\mathcal{P}_1,\mathcal{P}_2$ correspond to the support of minimum weight codewords of $\RM(k-r-1,k)$, which are simply $(r+1)$-dimensional affine spaces.

    Let $\pi:\mathbb{F}_2^k\rightarrow\mathbb{F}_2^{k-2}$ be the projection after removing the first two entries. Then $\pi:\mathcal{P}\rightarrow \mathbb{F}_2^{k-2}$ has to be injective and $0\notin \pi(\mathcal{P})$, otherwise $C_0$ would have repeated columns and the minimum distance of the quantum code would be $2$ or $C_0$ would be degenerate. 
    Now, since the intersection cannot be empty, let us take $v_o\in \mathcal{P}_1\cap \mathcal{P}_2$ and let $S_i=v_o+\mathcal{P}_i$.    
    Observe that if $v\in S_1\cap S_2$, then $\mathcal{P}_1\Delta \mathcal{P}_2=v+\mathcal{P}_1\Delta \mathcal{P}_2$. The previous implies that $S_1\cap S_2=\{0\}$, otherwise for $v\in S_1\cap S_2\setminus\{0\}$, $e_1+v,e_2+v\in \mathcal{P}_1\Delta \mathcal{P}_2$ and then $\pi$ was not injective on $\mathcal{P}$ or $v\in \{e_1,e_2,e_1+e_2\}$. If $v=e_i$, then $0,e_1+e_2\in \mathcal{P}_1\Delta \mathcal{P}_2$, a contradiction, since $\mathcal{P}$ cannot induced a degenerated code, then $v=e_1+e_2$ and for any $w\in \mathcal{P}_1\Delta \mathcal{P}_2\setminus\{e_1,e_2\}$, $\pi(w)=\pi(w+v)$, a contradiction. Thus $\mathcal{P}_1\cap \mathcal{P}_2=v_o$ and
    
    \begin{align*}
    |\mathcal{P}|=&|\mathcal{P}_1\Delta \mathcal{P}_2|\\
    =&|\mathcal{P}_1|+|\mathcal{P}_2|-2|\mathcal{P}_1\cap \mathcal{P}_2|\\=&2^{r+2}-2,\end{align*}
concluding the proof.\hfill$\square$
\end{proof}

The next family realizes these minimal $r$-orthogonal codes of dimension $2$.

\begin{theorem}\label{r-orth exp}
There exists a family of $r$-orthogonal codes with parameters
\[
\llbracket  2(2^{r+1}-2),\, 2,\, 3\rrbracket.
\]
\end{theorem}

\begin{proof}
Take $$A=\begin{pmatrix} \mathbf{1}\\ H_r^\ast\end{pmatrix}$$ and the codes $$C_1=\langle (1,0,0,0),(0,0,1,0),(0,1,0,1)\rangle,$$
$$C_2=\langle (0,0,1,1), (1,1,0,0)\rangle.$$ 
The code $D=[C_1\ C_2\ \cdots C_2]A$ has length $2^{r+2}-4$ and distance at least $2$. Since a generator matrix of the hull of $D$ is
$$\begin{pmatrix}
0&\mathbf{1}&0&\mathbf{1}\\
0&0&H_r^\ast&H_r^\ast\\
H_r^\ast&H_r^\ast&0&0\end{pmatrix}$$
and has no repeated columns, the distance of the code is at least $3$. By Corollary \ref{cor:incr}, the code $D$ is $r$-orthogonal.\hfill$\square$
\end{proof}

\begin{remark}
When $r=3$, the above family gives the smallest triorthogonal code with this property, which was previously found through a search method \cite{nezami2022classification}. 
\end{remark}

Taking affine spaces $\mathcal{P}_i$ such that $e_i\in \mathcal{P}_i$ and for each $(i,j)$ with $i\neq j$, $|\mathcal{P}_i\cap \mathcal{P}_j|=1$, induces a family of $k$ dimensional $r$-orthogonal codes of distance $3$. 

\begin{theorem}
There exists a family of $r$-orthogonal codes with parameters $[[n,k,3]]$, where 
\[
n=\begin{cases}
2k(2^r-1) & \ \text{for even} \ k \\  
2k(2^r-1)+1 & \ \text{for odd} \ k 
\end{cases}.
\]
\end{theorem}

\begin{proof}
    For $1\leq i\leq k$, let $X_i$ be the following affine space:
    $$\mathbf{1}+\langle \mathbf{1}+e_i, e_{k+(i-1)r+1},\ldots,e_{k+ir}\rangle\subseteq\mathbb{F}_2^{r(k+1)}.$$

    For $i\neq j$, we have $X_i\cap X_j=\{\mathbf{1}\}$, thus $$\mathcal{P}:=\Delta_{i=1}^k \mathcal{P}_i=\begin{cases}
    \bigcup_{i=1}^k \mathcal{P}_i& k=1\mod 2\\
    \bigcup_{i=1}^k \mathcal{P}_i\setminus\{\mathbf{1}\}&k=0\mod 2\end{cases},$$
    and
    $$\left|\mathcal{P}\right|=2^{r+1}k-k+\delta,$$
 where $\delta=0$ if $k$ is even and $\delta=1$ otherwise. After removing the elements $e_1,\ldots,e_k$ from $\Delta_{i=1}^k \mathcal{P}_i$, we obtain the desired length. To prove that the distance is $3$, it is enough to observe that $\pi(\mathcal{P})$, which is the projection into the last $r(k+1)-k$ coordinates, is injective. \hfill$\square$ 
\end{proof}

\begin{remark}
  We can view these codes as the direct sum of the combination of the code of dimension $2$ mentioned above, i.e., letting $A=\begin{pmatrix} \mathbf{1}\\ H_r^\ast\end{pmatrix}$, $C_1=\langle (1,0,0,0),(0,0,1,0),(0,1,0,1)\rangle$ and $C_2=\langle (0,0,1,1), (1,1,0,0)\rangle$,
  $$D_2=[C_1\ C_2\ \cdots\ C_2] A,$$
  and $D_1=[{C'}_1\ {C'}_2\ \cdots\ {C'}_2]A$, where ${C'}_1$ and ${C'}_2$ are the punctured codes from $C_1$ and $C_2$ in the third position. 

  The first code induces an $r$-orthogonal code of dimension $2$ and length $2(2^{r+1}-2)$, while the second code induces one of dimension $1$ and length $2^{r+1}-1$. To get dimension $k=2k'+t$ we need to add $k'$ copies of $D_2$ and $t$ copies of $D_1$. In the language of MP codes,
  $$[D_2\ \cdots\ D_2\ D_1]I_k.$$
  Later, we will construct families with better rates. 
\end{remark}

\begin{remark}
    The spaces of the form $\mathcal{P}=\mathbf{1}+\langle \mathbf{1}+e_{j_1},e_{j_2},\cdots e_{j_s}\rangle\subseteq\mathbb{F}_2^m$ correspond to the support of the evaluation of 
    $$f_\mathcal{P}=x_{j_1}\prod_{h\notin\{j_0,j_1,j_2,\ldots,j_s\}}(x_{j_0}+x_h+1),$$
where $j_0=\min [m]\setminus\{j_1,\ldots,j_s\}$. We can replace $j_0$ with any other element outside $\{j_1,\ldots,j_s\}$ and modify the product accordingly. Thus, instead of speaking of the space $\mathcal{P}$, we can just speak about the support of the evaluation of $f_\mathcal{P}$ and, equivalently, about its non-zeros. The intersection of two spaces $\mathcal{P}$ and $\mathcal{P}'$ is the non-zeros of the least common multiple of $f_\mathcal{P}$ and $f_{\mathcal{P}'}$. This is another way to see that in the previous result, the intersection of $\mathcal{P}_i$ and $\mathcal{P}_j$ has cardinality $1$. 
\end{remark}

Thus, controlling the intersections and the bijectivity of the projection are the challenges in describing the codes induced by RM codewords. If we sacrifice the latter, for example, we can create other families of codes with higher orthogonality. 

\begin{lemma}
    Let $k$ be a positive integer and let $R=\mathbb{F}_2[x_1,\ldots,x_{k+r+1}]$. Define the following polynomials:
    \begin{align*}
    f_1=&x_1(x_2+x_{k+r+1}+1)\prod_{j=3}^k (x_2+x_j+1)\\
    f_i=&x_i(x_1+x_{k+r}+1)\prod_{\substack{j=2\\ j\neq i}}^k (x_1+x_j+1)&2\leq i\leq k
    \end{align*}
    Let $f=\sum_{i=1}^k f_i$. Let $\mathcal{P}=\{P\in\mathbb{F}_2^{k+r+1}\ :\ f(P)=1,\ P\notin\{e_1,\ldots,e_k\}\}$. Then $\mathcal{P}$ induces an $r$-orthogonal code of length $(2^{r+1}-2^r-1)k+2^r$, dimension $k$ and distance $\geq 2$.  
\end{lemma}

\begin{proof}
    Observe that $f$ is a polynomial of degree $k$, so the vector $(f(P))_{P\in\mathbb{F}_2^{k+r+1}}$ defines a codeword dual to $RM(r,k+r+1)$, i.e., $\mathcal{P}$ is an $r$-orthogonal code. The dimension follows from the fact that we remove exactly the elements $e_1,\ldots,e_k$ from the non-zeros of $f$. On the other hand, $f_1$ defines the space $\mathcal{P}_1=\mathbf{1}+\langle\mathbf{1}+e_1,e_{k+1},\ldots,e_{k+r}\rangle$ and the rest of the $f_i$ define the spaces
    $$\mathcal{P}_i=\mathbf{1}+\langle\mathbf{1}+e_i,e_{k+1},\ldots,e_{k+r-1},e_{k+r+1}\rangle.$$

    Thus, $|\mathcal{P}_{i_1}\cap \mathcal{P}_{i_2}\cap\cdots\cap \mathcal{P}_{i_s}|=2^r$ for any $s\geq 2$ and $i_1>1$ and $|\mathcal{P}_1\cap \mathcal{P}_{i_1}\cap\cdots\cap \mathcal{P}_{i_s}|=2^{r-1}$ for any $s\geq 1$. Thus
    \begin{align*}
    |\mathcal{P}|=&(2^{r+1})k-k-2\left((k-1)2^{r-1}+{k-1\choose 2}2^ r\right)\\&+4\left({k-1\choose 2}2^{r-1}+{k-1\choose 3}2^r\right)-\cdots\\
    =&2^{r+1}k-k-2^r(k-1)=(2^{r+1}-2^r-1)k+2^r.
    \end{align*}
    The distance follows from the fact that none of the spaces $\mathcal{P}_i$ intersects the space $\langle e_1,\ldots,e_k\rangle$.\hfill$\square$
\end{proof}

Next, we develop a family of quantum CSS codes that are $(r+1)$-orthogonal for each $r \ge 1$. 
These codes are important in realizing logical $Z$-rotations by the angle $\frac{\pi}{2^{r}}$ for each $r \ge 1$. 
The parameters of such a family are 
\[
[[2^{r+1}+(2^{r}-1)k,k,2]]
\]
for each even integer $k\ge 0$. 
When $r=1$ and $r=2$ ($2$-orthogonal and triorthogonal), one retrieves the families $[[k+4,k,2]]$ H codes of \cite{jones2013multilevel} and $[[8+3k,k,2]]$ triorthogonal codes of \cite{bravyi2012magic}.

\begin{theorem}\label{T:4orthogonal}
There exists a family of $(r+1)$-orthogonal quantum CSS codes with parameters $[[(2^{r+1}-2^{r}-1)k+2^{r+1},k,2]]$ for each even integer $k$.   
\end{theorem}

\begin{proof}
    Take the $\frac{k}{2}$-dimensional $r$-orthogonal code defined in the previous lemma and let $A=\begin{pmatrix} A_1\\ A_0\end{pmatrix}$ be its corresponding $r$-orthogonal matrix. Define the code
    $$D=[C_1\ C_2\ \cdots\ C_2]A,$$
where $C_1=\langle (1,0),(0,1)\rangle$ and $C_2=\langle (1,1)\rangle$. Then $D$ is a code of length $(2^{r+1}-2^r-1)k+2^{r+1}$, dimension $k$ and distance $2$. It remains to see that it is $(r+1)$-orthogonal. So consider once again the points $\mathcal{P}$ defining the columns of $A$. Observe that the set $\{P\in \mathcal{P}\ :\ P_i=1\}$ for $1\leq i\leq k$ has to be equal to $\Delta_{i=1}^k \mathcal{P}'_i$, where $\mathcal{P}'_i=\mathbf{1}+\langle e_{k+1},\ldots,e_{k+r-1},e_{k+r+t}
    \rangle$, where $t=0$ if $i=1$ and $t=1$ otherwise. This set has cardinality divisible by $2$ since it is an affine space of dimension at least $1$. This is saying that any factor $u$ of $x_1\cdots x_{k+r+1}$ that is not divisible by $x_{k+1}\cdots x_{k+r+1}$ satisfies that $|\{P\in \mathcal{P}\ :\ u(P)=1\}|=0\mod 2$. Thus, we have $(r+1)$-orthogonality conditions.  \hfill$\square$
\end{proof}

In Appendix \ref{A:generalfamily}, we give an alternative proof with more details about the above family using the notation of \cite{bravyi2012magic}. 
In particular, we show that these codes do not require a correction operator, and we calculate the error probability per output magic state. 

\section{Application to triorthogonal codes with $\gamma \to 1$}\label{sec:gamma}
In this section, we construct a family of triorthogonal codes leading to $\gamma \to 1$. 
Previously, an instance of a quantum triorthogonal family satisfying $\gamma \to 1$ was provided in \cite{haah2018codes} via a randomized construction.

One main feature of the construction presented in Corollary \ref{C:newMPC1}, when restricted to $r=3$, is that the rate of the output quantum code $Q$ depends only on the length of the ingredient self-orthogonal code. 
In particular, the code $Q$ has rate 
\[
\frac{n_1}{n_1n_2}=\frac{1}{n_2}. 
\]
Furthermore, if the ingredient code $Q_1$ is also non-degenerate and $d_{Q_1}\ge d_{Q_2}$, then the minimum distance of $Q$ in Corollary \ref{C:newMPC1} is the same as the minimum distance of the quantum doubly-orthogonal ingredient code $Q_2$.

For now, suppose that there exists an infinite family of non-degenerate triorthogonal codes $[[n_i,1,d_i]]$ with increasing distance $d_i$ (we will prove the existence of such a family later).
Let $[[n_j,1,d_j]]$ be an infinite family of doubly-orthogonal (equivalently self-orthogonal) codes with 
\[
\lim_{j\to \infty}\frac{d_j}{n_j}=\delta>0.
\]
Then, applying the result of Corollary \ref{C:newMPC1} to pairs of $[[n_i',1,d_i']]$ and $[[n_i,1,d_i]]$ (assuming that the pairs have the property $d_i'\ge d_i$) implies the existence of an infinite family of triorthogonal codes with parameters 
\[
[[n_in_i',n_i',d_i]],
\]
which satisfies that $\lim_{i \to \infty}\frac{n_i'd_i}{n_in_i'}> 0$.
Moreover, this family satisfies 
\[
\gamma
=
\log_{d_i}\!\left(\frac{n_in_i'}{n_i'}\right)=\log_{d_i}\!\left(n_i\right).
\]
Furthermore, we have
\[
\begin{split}
\lim_{i\to \infty}\gamma_i&=\lim_{i\to \infty}(\log_{d_i}\!\left(n_i\right))\\&=
\lim_{i\to \infty}(\log_{\delta n_i}\!\left(n_i\right))=
\lim_{i\to \infty} \left(\frac{\ln(n_i)}{\ln(n_i)+\ln(\delta)}\right)\\&=1.
\end{split}
\]
Hence, this family reaches the constant cost function of $\gamma\to1$. 
To conclude, we only need to show the existence of
non-degenerate triorthogonal codes $[[n_i,1,d_i]]$ with increasing distance $d_i$ and 
self-orthogonal $[[n_j,1,d_j]]$ codes with 
$\lim_{j\to \infty}\frac{d_j}{n_j}=\delta>0.$

First, note that there is an extensive literature on self-dual codes; see, for example, \cite{rains2002self} and the references therein. 
Restricting the attention to binary linear codes, the existence of asymptotically good binary linear self-dual codes, both singly even and doubly even, was established in \cite{macwilliams1972good,thompson1973weighted}; see also \cite[Corollary 2 and Theorem 38]{rains2002self} for further discussion and additional details. 
Let $C$ be an $[n_i,\frac{n_i}{2},d_i]$ linear code from this family.
Then puncturing the code $C$ at an arbitrary position gives a code $C_o$ with parameters $[n_i-1,\frac{n_i}{2},d\ge d_i-1]$.
Shortening $C$ at the same position gives a code $C_e \subseteq C_o=C_e^\bot$ with parameters $[n_i-1,\frac{n_i}{2}-1,d\ge d_i]$.
The CSS code obtained from $C_e\subseteq C_o$ is a quantum code with parameters $[[n_i-1,1,\ge d_i-1]]$.
Repeating this process for all codes in this asymptotically good self-dual family yields a family of quantum codes with non-vanishing relative distance.

In the next section, we employ the Reed--Muller codes to build families of quantum codes that are triorthogonal, non-degenerate, and have a growing minimum distance. 
In particular, when $r=3$, this was the only remaining ingredient for building codes that allow magic $\ket{T}$ state distillation with asymptotic cost value $\gamma\to1$. 

\subsubsection{Punctured Reed--Muller codes and growing distance triorthogonal codes}

The following theorem gives a construction of triorthogonal quantum codes, based on punctured RM codes, with a growing distance. 
The idea of this result stems from the construction given in \cite{krishna2018towards}.
For any binary linear code $C$, we present the punctured and shortened code obtained from removing the coordinate $1$ of $C$ by $C^P$ and $C^S$. 
We refer the reader to Section \ref{sec: hamming and simplex} for the notation and a quick overview of Reed--Muller codes which we shall use below.
\begin{theorem}\label{T:RM1}
Let $C=\operatorname{RM}(s,m)$ with $m>3s$. 
The quantum CSS code of $C^S\subseteq C^P$ is triorthogonal with parameters 
\[
[[2^m-1,1,2^{s+1}-1]].
\]
This code is non-degenerate.
\end{theorem}

\begin{proof}
First, note that 
$C^\bot=\operatorname{RM}(m-s-1,m)$ and as $m>3s$, we have $s< m-s-1$ which implies that $C\subseteq C^\bot$.
Moreover, note that 
$C^{\ast2}=\operatorname{RM}(2s,m)$ and as $2s\le m-s-1$, we have $C^{\ast2} \subseteq C^\bot$. 

Next, consider the codes $C^S$ and $C^P$. 
They satisfy the following properties: 
\begin{itemize}
    \item $C^P=C^S\oplus \{{\bf 0,1}\}$;
    \item $C^S$ is self-orthogonal;
    \item $C^P$ has minimum distance $2^{m-s}-1$;
    \item $(C^S)^\bot=(C^\bot)^P$ and 
    $(C^P)^\bot=(C^\bot)^S$ (see \cite[Sections 1.5.1 and 1.5.3]{huffman2010fundamentals} for more details);
    \item $(C^S)^\bot$ has minimum distance $2^{s+1}-1$, and $(C^P)^\bot$ has minimum distance $2^{s+1}$.
\end{itemize}
Using the above properties, one can see that the CSS code built from $C^S \subseteq C^P$ has parameters $[[2^m-1,1,2^{s+1}-1]]$. 
This code is non-degenerate, as 
$(C^P)^\bot$ has minimum distance $2^{s+1}$.
To show that the code is triorthogonal, we use the criteria given in \cite[Theorem 2.3]{camps2024algebraic}, namely, we show that $(C^P)^{\ast2} \subseteq (C^S)^\bot$.
An easy calculation using the above properties shows that 
\begin{equation}\label{equ:ps condition}
(C^P)^{\ast2}=(C^S)^{\ast2}+C^S+\{{\textbf{0,1}}\}
\end{equation}
Moreover, as $C \subseteq C^{\ast2}\subseteq C^\bot$, then its shortened code satisfies 
\[
C^S \subseteq (C^S)^{\ast2}\subseteq (C^S)^\bot.
\]
Note that $C^S$ has only even vectors and thus ${\textbf{1}}\in (C^S)^\bot$. Finally, the last inclusion implies Equation \eqref{equ:ps condition}, showing that the CSS code obtained from $C^S \subseteq C^P$ is a triorthogonal code. 
$\hfill \square$
\end{proof}

In particular, one can choose the codes $\operatorname{RM}(s,m)$ with 
\[
(s,m)\in \{(i,3i+1): i\ge 1\},
\]
which all satisfy the conditions of the above theorem. 

We emphasize that one can modify the statement of Theorem \ref{T:RM1} to construct self-orthogonal codes. 
In particular, the following statement can be proved using similar steps. 
\begin{proposition}\label{T:RM2}
Let $C=\operatorname{RM}(s,m)$ with $m>2s$. 
The quantum CSS code built from $C^S\subseteq C^P$ is self-orthogonal with parameters 
\[
[[2^m-1,1,2^{s+1}-1]].
\]  
\end{proposition}
Now, one can use the above results as tools for generating the main ingredients of Corollary \ref{C:newMPC1}. 
This way, one can produce many triorthogonal codes with more than one logical qubit.

\begin{theorem}
Let $m_1>3s$ and $m_2>2s$ for some fixed positive integer $s$. 
Then, there exists an infinite class of triorthogonal codes with parameters 
\[
[[(2^{m_1}-1)(2^{m_2}-1),2^{m_1}-1,2^{s+1}-1]]
\]
\end{theorem}
For instance, choosing $m_1=4$, $m_2=3$ and $s=1$ gives the triorthogonal code $[[105,15,3]]$. 

Other quantum triorthogonal codes with distance $d=3$ and $d=7$, constructed in a similar fashion, are presented in Tables \ref{tab:MPC-RM} and \ref{tab:MPC-RM2}.

\begin{table}[t]
\centering
\caption{Parameters of $d=7$ triorthogonal codes obtained from MP of triorthogonal $\operatorname{RM}(2,m)$ (or punctured)  with $m\ge 7$ codes with $[[23,1,7]]$ quantum Golay code. 
Here $N$ shows the circuit width of the magic factory, where $N=k+\dim(X\text{-stabilizer})$.}
\label{tab:MPC-RM2}
\begin{tabular}{lcr}
$n$ & $k$ & $N$  \\
\toprule
$23\times(2^m-1)$& \ $(2^m-1)$ \ & $(2^m-1)+11(\frac{m^2+m}{2})$ \\
$23\times2^m$&$2^m$& $2^m+11(\frac{m^2+m+2}{2})$\\
\hline
&Examples&\\
\hline
2921 & 127 & 435 \\
2944 & 128 & 447 \\
5865 & 255 & 651 \\
5888 & 256 & 663 \\
\bottomrule
\end{tabular}
\end{table}

\section{Triorthogonal codes from more general quantum MP codes}\label{sec:generalMPC}
In this section, we consider some applications of Proposition \ref{P:twotriorthogonal} to form triorthogonal codes from MP codes, where one of the matrices is not a triorthogonal matrix, but it contains two blocks of triorthogonal matrices. 
A similar argument applies to generalize the result to higher $r$-orthogonal codes for $r\ge 4$.

First, consider the all-even code of an even length $n$. 
The generator matrix of such a code can be decomposed into 
$\begin{bmatrix}
A \\
B
\end{bmatrix},
$
where
\[
A=\begin{bmatrix}
1&1&0&0&0&\cdots&0&0\\ 
0&0&1&1&0&\cdots&0&0\\
\vdots&\vdots&\vdots&\vdots&\vdots&\cdots&\vdots&\vdots\\
0&0&0&0&0&\cdots&1&1\\
\end{bmatrix}
\]
and 
\[
B=\begin{bmatrix}
0&1&1&0&0&\cdots&0&0&0\\ 
0&0&0&1&1&\cdots&0&0&0\\
\vdots&\vdots&\vdots&\vdots&\vdots&\cdots&\vdots&\vdots&\vdots\\
0&0&0&0&0&\cdots&1&1&0\\
\end{bmatrix}.
\]
Using this decomposition, and the notation of Proposition \ref{P:twotriorthogonal} and its proof, we have $|B^0|=2$ and $d(D^\bot)=n$. 
So we have the following result. 
\begin{corollary}
Let $Q$ be the quantum CSS code built from $C_2\subseteq C_1$, with parameters $[[n,1,d]]$ where $n$ is odd and $C_2$ is a self-orthogonal code with $C_1=C_2\oplus\{0_n,1_n\}$. 
Then, there exist quantum triorthogonal codes with parameters
\[
[[n(d-1),2,d-1]]
\]
and 
\[
[[n(d+1),2,d]].
\]
\end{corollary}

\begin{proof}
We apply the result of Proposition \ref{P:twotriorthogonal} to the code $Q$ and the all-even codes of length $d-1$ and $d+1$, respectively. 
As we mentioned above, the generator matrix of the all-even code can be decomposed into two triorthogonal sub-matrices $A$ and $B$ with $|B^0|=2$. 
Now, the MP construction of 
Proposition \ref{P:twotriorthogonal} gives the codes
\[
[[n(d-1),2,d-1]]
\]
and 
\[
[[n(d+1),2,d]].
\]
$\hfill \square$    
\end{proof}

In particular, applying the above corollary to the small-length quantum self-orthogonal codes presented in Table 1 of \cite{jain2025transversal} and \cite[Example III.5]{dastbasteh2026quantum} gives the quantum triorthogonal codes presented in the second and third columns of the table below. 

\begin{center}
\begin{tabular}{c|c|c}
 $Q$ & $d-1$ type  & $d+1$ type \\\hline
 $[[7,1,3]]$   & $[[14,2,2]]$   & $[[28,2,3]]$   \\
 $[[17,1,5]]$  & $[[68,2,4]]$ & $[[102,2,5]]$ \\
 $[[23,1,7]]$  & $[[138,2,6]]$& $[[184,2,7]]$ \\
 $[[41,1,9]]$  & $[[328,2,8]]$  & $[[410,2,9]]$  \\
 $[[47,1,11]]$ & $[[470,2,10]]$  & $[[564,2,11]]$ 
\end{tabular}
\end{center}

So far we mainly discussed a special case of Proposition \ref{P:twotriorthogonal} by restricting to only all-even codes. 
In general, Proposition \ref{P:twotriorthogonal} allows us to construct many other codes with a larger number of logical qubits as long as suitable block-triorthogonal matrices are designed. 
Therefore, we discuss a natural approach for forming matrices that consist of two even-triorthogonal row
blocks.  

Let $C_1$ and $C_2$ be two binary linear codes with parameters $[n,k_1,d_1]$ and $[n,k_2,d_2]$. The Plotkin sum of $C_1$ and $C_2$ has generator matrix 
\[
\begin{bmatrix}
G_1 & G_1\\
0 & G_2
\end{bmatrix},
\]
where $G_1$ and $G_2$ are generator matrices of $C_1$ and $C_2$, respectively. We represent such a code by $P(C_1,C_2)$; it has parameters 
\[
[2n,k_1+k_2,\min\{2d_1,d_2\}].\] 
Moreover, the code $P(C_1,C_2)^\bot$ has parameters 
\[
[2n,2n-k_1-k_2,\min\{d_1^\bot,2d_2^\bot\}],\] 
where $d_1^\bot=d(C_1^\bot)$ and $d_2^\bot=d(C_2^\bot)$.

One feature of the Plotkin sum is the following. The proof is easy, and we omit it here.

\begin{proposition}\label{P:plotkin-two-triorthogonal}
Let $C_2$ be a binary linear code with an even triorthogonal generator matrix, and let $C_1$ be an arbitrary linear code of the same length. Then $P(C_1,C_2)$ has a generator matrix that consists of two triorthogonal block matrices.  
\end{proposition}

Next, we use the above result to build quantum triorthogonal codes with larger minimum distances. 

\begin{theorem}\label{T:twoorthogonal2}
Let $Q$ be a quantum self-orthogonal CSS code corresponding to $C_2\subseteq C_2\oplus\{0_n,1_n\}$ with parameters
$[[n,1,d>1]]$, where $n$ is odd.
Let $C$ be a binary linear code of length $m$ with generator matrix $G_1$, and let $G_2$ be an even triorthogonal matrix generating a binary linear code $C'$. 
Then, there exists a binary triorthogonal quantum code with parameters $[[2nm,k,\ge d_z]]$, where
\[
d_z=\min\{d,d(C^\bot),2d(C'^\bot)\}
\]
and 
\[
k=
\begin{cases}
m+1 & m \ \text{odd}\\
m & m \ \text{even}.
\end{cases}
\]
\end{theorem}

\begin{proof}
Let
\[
G=\begin{bmatrix}
G_1 & G_1\\
0   & G_2
\end{bmatrix}.
\]
Then $G$ is a generator matrix of $P(C,C')$ that consists of two blocks of triorthogonal matrices by the previous proposition.
Moreover, $P(C,C')^\bot$ has minimum distance $\min\{d(C^\bot),2d(C'^\bot)\}$.
Now, applying Proposition \ref{P:twotriorthogonal} implies a quantum triorthogonal code with parameters 
\[
[[2nm,m,\ge \min\{d,d(C^\bot),2d(C'^\bot)\}]].
\]
The $X$-stabilizers of such codes are 
\[
M=\begin{bmatrix}
\underbrace{C_2, \dots, C_2}_{\text{\# rows of } G_1}, & \underbrace{\overline{1}_n, \dots, \overline{1}_n}_{\text{\# rows of } G_2}
\end{bmatrix}
G
\]
and the $m$ $X$-logical operators are $(e_i,0_{nm})$ for each $1\le i \le m$, where $e_i$ has all ones in the $i$-th block and zeros in the remaining $m-1$ blocks.
If $m$ is odd, then the vector $e=(0_{nm},1_{nm})$ is also triorthogonal to all the mentioned logical operators and has the odd weight $nm$. 
Finally, $e\ast(m_1,m_2)\ast(m_1',m_2')=m_2\ast m_2'$ has an even weight, for each two $(m_1,m_2)$ and $(m_1',m_2')$ rows of $M$ as the second $nm$ column block of $M$ has orthogonal rows. 
This proves the existence of a quantum triorthogonal code with parameters 
\[
[[2nm,k,\ge \min\{d,d(C^\bot),2d(C'^\bot)\}]]
\]
and concludes the proof.\hfill$\square$
\end{proof}

\begin{example}
Let $Q$ be the $[[15,1,3]]$ quantum triorthogonal code with the $X$-stabilizer corresponding to the binary linear code $C'$. Let $C$ be the all-even code of length $15$. 
Then, the code $P(C,C')$ has a generator matrix 
\[
G
=\begin{bmatrix}
G_1 & G_1\\
0   & G_2
\end{bmatrix},
\]
where $G_1$ and $G_2$ are generator matrices of $C$ and $C'$, respectively. 
Note that $P(C,C')^\bot$ has minimum distance $d=6$. 

Now, applying the result of Theorem \ref{T:twoorthogonal2}, using the quantum Golay code $[[23,1,7]]$, implies the existence of a $[[690,16,6]]$ quantum triorthogonal code.
\end{example}

\begin{example}
Let $C=\F_2^{16}$ or any other binary linear code satisfying $d(C^\bot)\ge 8$,  and $C'=\operatorname{RM}(1 ,4)$. 
In this case, $P(C,C')^\bot$ has minimum distance $2d(\operatorname{RM}(1 ,4)^\bot)=8$. 
Applying the result of Theorem \ref{T:twoorthogonal2}, using the quantum Golay code $[[23,1,7]]$, gives a quantum triorthogonal code with parameters $[[736,16,7]]$.
Note that one can substitute the code $C$ with a smaller-dimensional code to get another quantum triorthogonal code with the same parameters, but a smaller magic factory width $N$.
\end{example}

One can generalize the above examples to an infinite class of quantum triorthogonal codes as follows.

\begin{corollary}\label{C:matrix-plotkin}
Let $m\ge 0$ and $s\ge 0$ be integers such that $m>3s$. Let $Q$ be a quantum self-orthogonal CSS code corresponding to $C_2\subseteq C_2\oplus\{0_n,1_n\}$ with parameters
$[[n,1,d]]$, where $n$ is odd.
Then, there exists a quantum triorthogonal code with parameters
\[
[[n2^{m+1},2^m,\min\{d,2^{s+2}\}]].
\]
Furthermore, such a code has factory width
\[
2^m+\dim(\operatorname{RM}(s+1 ,m))\dim(C_2)+\dim(\operatorname{RM}(s ,m)).
\]
\end{corollary}

\begin{proof}
The proof follows from the idea of the previous example by choosing $C'=\operatorname{RM}(s ,m)$ and $C=\operatorname{RM}(s+1 ,m)$. 
Note also that $P(C,C')=\operatorname{RM}(s+1,m+1)$. 
Hence $P(C,C')^\perp=\operatorname{RM}(m-s-1,m+1)$, whose minimum distance is $2^{s+2}$. 
Its minimum-weight codewords are the indicator vectors of the $(s+2)$-dimensional affine subspaces of $\mathbb F_2^{m+1}$, and some of these meet the first half of the coordinates. 
By Proposition \ref{P:twotriorthogonal}, the minimum distance is therefore exactly $\min\{d,2^{s+2}\}$.

The $X$-stabilizer group of such quantum code has dimension $\dim(\operatorname{RM}(s+1 ,m))\dim(C_2)+\dim(\operatorname{RM}(s ,m))$. 
Thus, the factory width is the claimed value. \hfill$\square$
\end{proof}
We emphasize that, for the codes generated by the above corollary,  
the overhead exponent 
\[
\gamma=\frac{\log(2n)}{\log(\min\{d,2^{s+2}\})}\]
does not depend on $m$.

\begin{example}
Let $s=1$ and $m=4$. 

We choose $Q$ to be the $[[17,1,5]]$ quantum self-orthogonal code. 
Applying Corollary \ref{C:matrix-plotkin} gives a quantum triorthogonal code with parameters $[[544,16,5]]$.

We choose $Q$ to be the $[[23,1,7]]$  quantum Golay code. 
Applying Corollary \ref{C:matrix-plotkin} implies a quantum triorthogonal code with parameters $[[736,16,7]]$.
\end{example}
Other instances of quantum codes generated this way are presented in Table \ref{tab:plotkin}.

Note also that one can use the Plotkin sum to build a variety of matrices decomposable into two blocks of triorthogonal codes. 
We give further discussion of this in Appendix \ref{A: two blocks}.

\begin{table}[t]
\caption{Instances of Corollary \ref{C:matrix-plotkin}.}
\label{tab:plotkin}
\centering
\footnotesize
\setlength{\tabcolsep}{4.5pt}
\begin{tabular}{@{}ccllrrr@{}}
\toprule
$s$ & $m$ & Inner code $Q$ & Final code & $N$\\
\midrule
0 & 1 & $[[7,1,3]]$ & $[[28,2,3]]$ & 9 \\
0 & 2 & $[[7,1,3]]$ & $[[56,4,3]]$ & 14 \\
0 & 4 & $[[7,1,3]]$ & $[[224,16,3]]$ & 32\\
0 & 1 & $[[17,1,5]]$ & $[[68,2,4]]$ & 19 \\
\midrule
1 & 4 & $[[17,1,5]]$ & $[[544,16,5]]$ & 109 \\
1 & 5 & $[[17,1,5]]$ & $[[1088,32,5]]$ & 166 \\
1 & 6 & $[[17,1,5]]$ & $[[2176,64,5]]$ & 247 \\
1 & 4 & $[[23,1,7]]$ & $[[736,16,7]]$ & 142 \\
1 & 5 & $[[23,1,7]]$ & $[[1472,32,7]]$ & 214 \\
1 & 6 & $[[23,1,7]]$ & $[[2944,64,7]]$ & 313 \\
\bottomrule
\end{tabular}
\end{table}

\section{Conclusion}

We introduced matrix product codes as a systematic way to construct
$r$-orthogonal codes, which realize a transversal logical $Z$-rotation by
$\pi/2^{r-1}$. The construction takes seed codes from two consecutive levels
of the Clifford hierarchy and returns a code at the higher level. In
particular, an $r$-orthogonal $[[n_1,1,d_1]]$ code and an $(r-1)$-orthogonal
$[[n_2,1,d_2]]$ code of odd lengths give an $r$-orthogonal code of length
$n_1n_2$ with $n_1$ logical qubits. The rate $1/n_2$ depends only on the
lower-level seed, and strong orthogonality is inherited from the seeds, so the
logical gate requires no Clifford correction.

For example, the $[[105,15,3]]$ strongly triorthogonal code is obtained from
the $[[15,1,3]]$ Reed--Muller code and the Steane code. 
We also proposed a qubit recycling technique that exploits the structure of the MP codes. 
In particular, a recycling
schedule is found once for the small base code, and is then carried through
the matrix product. 
For the $[[105,15,3]]$ code, this reduces the footprint
of the factory from $27$ to $21$ logical qubits for $15$ output $T$ states,
and the same procedure applies to every factory built from our MP
construction.

The MP framework also contains codes that were previously known only as isolated
examples. We showed that minimal $r$-orthogonal codes correspond to supports
of Reed--Muller codewords. From this correspondence, we obtained the shortest
$r$-orthogonal codes with two logical qubits and distance $3$, together
with distance-three codes for any number of logical qubits. At distance $2$,
the family $[[2^{r}+(2^{r-1}-1)k,\,k,\,2]]$ with even $k$, places the H codes
\cite{jones2013multilevel} and the Bravyi--Haah codes \cite{bravyi2012magic} in a single
framework. The codes in this family require no correction operator and
reduce an input error rate $p$ to $\bigl((2^{r-1}-1)k+1\bigr)p^{2}$ per
output, to leading order.

Using the MP construction, we also obtained the following families of
triorthogonal codes:
\begin{itemize}
    \item $[[(2^{m_1}-1)(2^{m_2}-1),\,2^{m_1}-1,\,2^{s+1}-1]]$ codes for
    $m_1>3s$ and $m_2>2s$, from punctured and shortened Reed--Muller codes;
    \item $[[n(d-1),2,d-1]]$ and $[[n(d+1),2,d]]$ codes, from any
    self-orthogonal $[[n,1,d]]$ code of odd length $n$;
    \item $[[n2^{m+1},\,2^{m},\,\min\{d,2^{s+2}\}]]$ codes for $m>3s$, from
    the same self-orthogonal codes combined with Reed--Muller codes.
\end{itemize}
We also showed that combining growing-distance triorthogonal seeds with
asymptotically good self-dual codes gives a triorthogonal family with
$\gamma\to 1$.

Our results suggest several directions for future research. First, it would
be interesting to know whether a generalization of the matrix product, in the
spirit of the lifted and balanced products for qLDPC codes, yields explicit,
structured finite-length families with $\gamma<1$. Secondly, our recycling
technique uses only the structure of the base code. Optimizing jointly over
the inner code and over the order of the rotations inside each block could
reduce the footprint further. Finally, we believe that, besides its
application to magic state distillation, the structure of our MP codes is
directly applicable to parallel magic state preparation via the
code-switching protocols of \cite{heussen2025efficient,dastbasteh2026quantum}.
We leave these problems for future work.

\section*{Acknowledgment}
RD and JEM acknowledge the support by the Spanish Ministry of Economy and Competitiveness through the MADDIE (Grant No. PID2022-137099NBC44) and GEMMA projects ``Generative AI and quantum approaches for advanced sensing and communications'' (Grant No. PID2025-169068NB-C33), and by the Ministry of Economic Affairs and Digital Transformation of the Spanish Government through the QUANTUM ENIA project call - QUANTUM SPAIN project, and by the European Union through the Recovery, Transformation and Resilience Plan - NextGenerationEU within the framework of the Digital Spain 2026 Agenda. 
RD and JEM also acknowledge support from the Basque Quantum (BasQ) strategy through the ``Decoders for quantum low-density parity-check codes'' (qLDPCDec) project. 

EB and ECM are supported by the grant ANR-22-CPJ2-004701. GZ is supported by Plan France 2030 through the Project NISQ2LSQ, ANR-22-PETQ-0006.

We acknowledge the use of AI-assisted tools (Anthropic models) for creating Figure 1 and for grammar and readability checks.

\bibliographystyle{IEEEtran}
\bibliography{Zrotation}

@inproceedings{allerton,
  title={Binary Triorthogonal and {CSS-T} codes for quantum error correction},
  author={Camps-Moreno, Eduardo and L{\'o}pez, Hiram H and Matthews, Gretchen L and Ruano, Diego and San--Jos{\'e}, Rodrigo and Soprunov, Ivan},
  booktitle={2024 60th Annual Allerton Conference on Communication, Control, and Computing},
  pages={01--06},
  year={2024},
  organization={IEEE}
}

@article{mpc,
  title={Matrix-product codes over $\mathbb{F}_q$},
  author={Blackmore, Tim and Norton, Graham H},
  journal={Applicable Algebra in Engineering, Communication and Computing},
  volume={12},
  number={6},
  pages={477--500},
  year={2001},
  publisher={Springer}
}

@Article{BBcodes,
author={Bravyi, Sergey
and Cross, Andrew W.
and Gambetta, Jay M.
and Maslov, Dmitri
and Rall, Patrick
and Yoder, Theodore J.},
title={High-threshold and low-overhead fault-tolerant quantum memory},
journal={Nature},
year={2024},
month={Mar},
day={01},
volume={627},
number={8005},
pages={778-782},
issn={1476-4687},
doi={10.1038/s41586-024-07107-7},
url={https://doi.org/10.1038/s41586-024-07107-7}
}

@Article{xu2023constantoverheadfaulttolerantquantumcomputation,
author={Xu, Qian
and Bonilla Ataides, J. Pablo
and Pattison, Christopher A.
and Raveendran, Nithin
and Bluvstein, Dolev
and Wurtz, Jonathan
and Vasi{\'{c}}, Bane
and Lukin, Mikhail D.
and Jiang, Liang
and Zhou, Hengyun},
title={Constant-overhead fault-tolerant quantum computation with reconfigurable atom arrays},
journal={Nature Physics},
year={2024},
month={Jul},
day={01},
volume={20},
number={7},
pages={1084-1090},
issn={1745-2481},
doi={10.1038/s41567-024-02479-z},
url={https://doi.org/10.1038/s41567-024-02479-z}
}

@ARTICLE{kasaicodes,
       author = {{Zhao}, Chen and {Duckering}, Casey and {Gu}, Andi and {Maskara}, Nishad and {Zhou}, Hengyun},
        title = {Towards Ultra-High-Rate Quantum Error Correction with Reconfigurable Atom Arrays},
      journal = {arXiv e-prints},
         year = 2026,
        month = apr,
          eid = {arXiv:2604.16209},
        pages = {arXiv:2604.16209},
          doi = {10.48550/arXiv.2604.16209},
archivePrefix = {arXiv},
       eprint = {2604.16209},
 primaryClass = {quant-ph},
       adsurl = {https://ui.adsabs.harvard.edu/abs/2026arXiv260416209Z}
}

@article{eastinknill,
  title = {Restrictions on Transversal Encoded Quantum Gate Sets},
  author = {Eastin, Bryan and Knill, Emanuel},
  journal = {Phys. Rev. Lett.},
  volume = {102},
  issue = {11},
  pages = {110502},
  numpages = {4},
  year = {2009},
  month = {Mar},
  publisher = {American Physical Society},
  doi = {10.1103/PhysRevLett.102.110502},
  url = {https://link.aps.org/doi/10.1103/PhysRevLett.102.110502}
}

@article{megaquop,
author = {Preskill, John},
title = {Beyond {NISQ}: The Megaquop Machine},
year = {2025},
issue_date = {September 2025},
publisher = {Association for Computing Machinery},
address = {New York, NY, USA},
volume = {6},
number = {3},
url = {https://doi.org/10.1145/3723153},
doi = {10.1145/3723153},
journal = {ACM Transactions on Quantum Computing},
month = apr,
articleno = {18},
numpages = {7}
}

@article{diego,
  title={Construction and decoding of matrix-product codes from nested codes},
  author={Hernando, Fernando and Lally, Kristine and Ruano, Diego},
  journal={Applicable Algebra in Engineering, Communication and Computing},
  volume={20},
  number={5},
  pages={497--507},
  year={2009},
  publisher={Springer}
}

@article{kasami,
  title={On the weight structure of {Reed-Muller} codes},
  author={Kasami, Tadao and Tokura, Nobuki},
  journal={IEEE Transactions on Information Theory},
  volume={16},
  number={6},
  pages={752--759},
  year={1970},
  publisher={IEEE}
}

@article{litinski,
  doi = {10.22331/q-2019-12-02-205},
  url = {https://doi.org/10.22331/q-2019-12-02-205},
  title = {Magic {S}tate {D}istillation: not as costly as you think},
  author = {Litinski, Daniel},
  journal = {{Quantum}},
  issn = {2521-327X},
  publisher = {{Verein zur F{\"{o}}rderung des Open Access Publizierens in den Quantenwissenschaften}},
  volume = {3},
  pages = {205},
  month = dec,
  year = {2019}
}

@ARTICLE{recycling,
       author = {{Xu}, Shifan and {Liu}, Kun and {Rall}, Patrick and {He}, Zhiyang and {Ding}, Yongshan},
        title = "{Distilling Magic States in the Bicycle Architecture}",
      journal = {arXiv e-prints},
         year = 2026,
        month = feb,
          eid = {arXiv:2602.20546},
        pages = {arXiv:2602.20546},
          doi = {10.48550/arXiv.2602.20546},
archivePrefix = {arXiv},
       eprint = {2602.20546},
 primaryClass = {quant-ph},
       adsurl = {https://ui.adsabs.harvard.edu/abs/2026arXiv260220546X}
}

@ARTICLE{degen,
  author={Fuentes, Patricio and {Etxezarreta Martinez}, Josu and Crespo, Pedro M. and Garcia-Frías, Javier},
  journal={IEEE Access}, 
  title={Degeneracy and Its Impact on the Decoding of Sparse Quantum Codes}, 
  year={2021},
  volume={9},
  number={},
  pages={89093-89119},
  doi={10.1109/ACCESS.2021.3089829},
  url = {https://ieeexplore.ieee.org/abstract/document/9456887}}

@ARTICLE{cultivation,
       author = {{Gidney}, Craig and {Shutty}, Noah and {Jones}, Cody},
        title = {Magic state cultivation: growing {T} states as cheap as {CNOT} gates},
      journal = {arXiv e-prints},
         year = 2024,
        month = sep,
          eid = {arXiv:2409.17595},
        pages = {arXiv:2409.17595},
          doi = {10.48550/arXiv.2409.17595},
archivePrefix = {arXiv},
       eprint = {2409.17595},
 primaryClass = {quant-ph},
       adsurl = {https://ui.adsabs.harvard.edu/abs/2024arXiv240917595G}
}

@article{distillation,
  title = {Universal quantum computation with ideal {C}lifford gates and noisy ancillas},
  author = {Bravyi, Sergey and Kitaev, Alexei},
  journal = {Phys. Rev. A},
  volume = {71},
  issue = {2},
  pages = {022316},
  numpages = {14},
  year = {2005},
  month = {Feb},
  publisher = {American Physical Society},
  doi = {10.1103/PhysRevA.71.022316},
  url = {https://link.aps.org/doi/10.1103/PhysRevA.71.022316}
}

@article{cultivation2,
  title = {High Rate Magic State Cultivation on the Surface Code},
  author = {Vaknin, Yotam and Jacoby, Shoham and Grimsmo, Arne and Retzker, Alex},
  journal = {PRX Quantum},
  volume = {7},
  issue = {1},
  pages = {010353},
  numpages = {20},
  year = {2026},
  month = {Mar},
  publisher = {American Physical Society},
  doi = {10.1103/p8tw-6kq9},
  url = {https://link.aps.org/doi/10.1103/p8tw-6kq9}
}

@Article{constantCCZ,
author={Wills, Adam
and Hsieh, Min-Hsiu
and Yamasaki, Hayata},
title={Constant-overhead magic state distillation},
journal={Nature Physics},
year={2025},
month={Nov},
day={01},
volume={21},
number={11},
pages={1842-1846},
issn={1745-2481},
doi={10.1038/s41567-025-03026-0},
url={https://doi.org/10.1038/s41567-025-03026-0}
}

@inproceedings{Golowich25,
author = {Golowich, Louis and Guruswami, Venkatesan},
title = {Asymptotically Good Quantum Codes with Transversal Non-{C}lifford Gates},
year = {2025},
isbn = {9798400715105},
publisher = {Association for Computing Machinery},
address = {New York, NY, USA},
url = {https://doi.org/10.1145/3717823.3718234},
doi = {10.1145/3717823.3718234},
booktitle = {Proceedings of the 57th Annual ACM Symposium on Theory of Computing},
pages = {707–717},
numpages = {11},
location = {Prague, Czechia},
series = {STOC '25}
}

@inproceedings{nguyen25,
author = {Nguyen, Quynh T.},
title = {Good Binary Quantum Codes with Transversal {CCZ} Gate},
year = {2025},
isbn = {9798400715105},
publisher = {Association for Computing Machinery},
address = {New York, NY, USA},
url = {https://doi.org/10.1145/3717823.3718186},
doi = {10.1145/3717823.3718186},
booktitle = {Proceedings of the 57th Annual ACM Symposium on Theory of Computing},
pages = {697–706},
numpages = {10},
location = {Prague, Czechia},
series = {STOC '25}
}

@ARTICLE{sanjose26,
       author = {{San-Jos{\'e}}, Rodrigo},
        title = "{Asymptotically good binary triorthogonal codes and higher-level transversal gates}",
      journal = {arXiv e-prints},
         year = 2026,
        month = sep,
          eid = {arXiv:2609.08203},
        pages = {arXiv:2609.08203},
          doi = {10.48550/arXiv.2609.08203},
archivePrefix = {arXiv},
       eprint = {2609.08203},
 primaryClass = {math.IT},
       adsurl = {https://ui.adsabs.harvard.edu/abs/2026arXiv260908203S}
}

@ARTICLE{willsAsymptotic,
       author = {{Wills}, Adam},
        title = {Improved Quantum Codes with Transversal {T} Gates},
      journal = {arXiv e-prints},
         year = 2026,
        month = aug,
          eid = {arXiv:2608.24000},
        pages = {arXiv:2608.24000},
          doi = {10.48550/arXiv.2608.24000},
archivePrefix = {arXiv},
       eprint = {2608.24000},
 primaryClass = {quant-ph},
       adsurl = {https://ui.adsabs.harvard.edu/abs/2026arXiv260824000W}
}

@ARTICLE{Li26,
       author = {{Li}, Yiming and {Li}, Zimu and {Han}, Zhengyi and {Liu}, Zi-Wen},
        title = "{Parallelizable and addressable transversal non-{C}lifford gates on good quantum LDPC codes}",
      journal = {arXiv e-prints},
         year = 2026,
        month = sep,
          eid = {arXiv:2609.26691},
        pages = {arXiv:2609.26691},
          doi = {10.48550/arXiv.2609.26691},
archivePrefix = {arXiv},
       eprint = {2609.26691},
 primaryClass = {quant-ph},
       adsurl = {https://ui.adsabs.harvard.edu/abs/2026arXiv260926691L}
}

@ARTICLE{Lin26,
       author = {{Lin}, Tongyin and {Wu}, Bujiao and {Cheng}, Bin},
        title = {Asymptotically Good Quantum Codes with Addressable Transversal {T} Gates},
      journal = {arXiv e-prints},
         year = 2026,
        month = sep,
          eid = {arXiv:2609.35129},
        pages = {arXiv:2609.35129},
          doi = {10.48550/arXiv.2609.35129},
archivePrefix = {arXiv},
       eprint = {2609.35129},
 primaryClass = {quant-ph},
       adsurl = {https://ui.adsabs.harvard.edu/abs/2026arXiv260935129L}
}

@ARTICLE{classificationTri,
       author = {{Wills}, Adam and {Jain}, Shubham P. and {Singh}, Shraddha},
        title = {Classification of Generalised Triorthogonal Codes through Length 54},
      journal = {arXiv e-prints},
         year = 2026,
        month = sep,
          eid = {arXiv:2609.30860},
        pages = {arXiv:2609.30860},
          doi = {10.48550/arXiv.2609.30860},
archivePrefix = {arXiv},
       eprint = {2609.30860},
 primaryClass = {quant-ph},
       adsurl = {https://ui.adsabs.harvard.edu/abs/2026arXiv260930860W}
}

@ARTICLE{AItri,
       author = {{Jain}, Shubham P. and {Wills}, Adam and {Singh}, Shraddha},
        title = "{Symmetry and {AI}-assisted discovery of magic-state factories}",
      journal = {arXiv e-prints},
         year = 2026,
        month = oct,
          eid = {arXiv:2610.06535},
        pages = {arXiv:2610.06535},
          doi = {10.48550/arXiv.2610.06535},
archivePrefix = {arXiv},
       eprint = {2610.06535},
 primaryClass = {quant-ph},
       adsurl = {https://ui.adsabs.harvard.edu/abs/2026arXiv261006535J}
}

@ARTICLE{maleableTri,
       author = {{Singh}, Shraddha and {Gidney}, Craig and {Jones}, Cody},
        title = "{Borrowed Identities: Malleable Distillation Factories and a Unified Numerical Search}",
      journal = {arXiv e-prints},
         year = 2026,
        month = jun,
          eid = {arXiv:2606.28518},
        pages = {arXiv:2606.28518},
          doi = {10.48550/arXiv.2606.28518},
archivePrefix = {arXiv},
       eprint = {2606.28518},
 primaryClass = {quant-ph},
       adsurl = {https://ui.adsabs.harvard.edu/abs/2026arXiv260628518S}
}

@ARTICLE{compactTri,
       author = {{Jacinto}, Hugo and {Valcarce}, Xavier and {Barizien}, Victor and {Gouzien}, {\'E}lie and {Sangouard}, Nicolas},
        title = "{Exploring the landscape of compact magic-state distillation factories}",
      journal = {arXiv e-prints},
         year = 2026,
        month = jun,
          eid = {arXiv:2606.07734},
        pages = {arXiv:2606.07734},
          doi = {10.48550/arXiv.2606.07734},
archivePrefix = {arXiv},
       eprint = {2606.07734},
 primaryClass = {quant-ph},
       adsurl = {https://ui.adsabs.harvard.edu/abs/2026arXiv260607734J}
}

@ARTICLE{QMPC,
       author = {{Jaramillo-Velez}, Delio and {Neri}, Alessandro and {Patra}, Adway and {Ruano}, Diego and {Salizzoni}, Flavio},
        title = {Quantum Matrix-Product Codes: {CSS-T} Characterization and Maximality},
      journal = {arXiv e-prints},
         year = 2026,
        month = sep,
          eid = {arXiv:2609.08520},
        pages = {arXiv:2609.08520},
          doi = {10.48550/arXiv.2609.08520},
archivePrefix = {arXiv},
       eprint = {2609.08520},
 primaryClass = {math.IT},
       adsurl = {https://ui.adsabs.harvard.edu/abs/2026arXiv260908520J}
}

@ARTICLE{magicsComp,
       author = {{Gao}, Jianshuo and {Yuan}, Xiao and {Yao}, Yuan},
        title = {A Resource Comparison of Logical {T}-State Preparation},
      journal = {arXiv e-prints},
         year = 2026,
        month = may,
          eid = {arXiv:2605.26522},
        pages = {arXiv:2605.26522},
          doi = {10.48550/arXiv.2605.26522},
archivePrefix = {arXiv},
       eprint = {2605.26522},
 primaryClass = {quant-ph},
       adsurl = {https://ui.adsabs.harvard.edu/abs/2026arXiv260526522G}
}

@article{
femoco,
author = {Markus Reiher  and Nathan Wiebe  and Krysta M. Svore  and Dave Wecker  and Matthias Troyer },
title = {Elucidating reaction mechanisms on quantum computers},
journal = {Proceedings of the National Academy of Sciences},
volume = {114},
number = {29},
pages = {7555-7560},
year = {2017},
doi = {10.1073/pnas.1619152114},
URL = {https://www.pnas.org/doi/abs/10.1073/pnas.1619152114},
eprint = {https://www.pnas.org/doi/pdf/10.1073/pnas.1619152114}}

@inproceedings{panteleev2022asymptotically,
  title={Asymptotically good quantum and locally testable classical {LDPC} codes},
  author={Panteleev, Pavel and Kalachev, Gleb},
  booktitle={Proceedings of the 54th annual ACM SIGACT symposium on theory of computing},
  pages={375--388},
  year={2022}
}

@article{tillich2013quantum,
  title={Quantum {LDPC} codes with positive rate and minimum distance proportional to the square root of the blocklength},
  author={Tillich, Jean-Pierre and Z{\'e}mor, Gilles},
  journal={IEEE Transactions on Information Theory},
  volume={60},
  number={2},
  pages={1193--1202},
  year={2013},
  publisher={IEEE}
}

@article{CS96,
  title={Good quantum error-correcting codes exist},
  author={Calderbank, A Robert and Shor, Peter W},
  journal={Physical Review A},
  volume={54},
  number={2},
  pages={1098--1105},
  year={1996},
  publisher={APS}
}

@article{steane1996multiple,
  title={Multiple-particle interference and quantum error correction},
  author={Steane, Andrew},
  journal={Proceedings of the Royal Society of London. Series A: Mathematical, Physical and Engineering Sciences},
  volume={452},
  number={1954},
  pages={2551--2577},
  year={1996},
  publisher={The Royal Society London}
}

@article{demarti2024decoding,
  title={Decoding algorithms for surface codes},
  author={deMarti iOlius, Antonio and Fuentes, Patricio and Or{\'u}s, Rom{\'a}n and Crespo, Pedro M and {Etxezarreta Martinez}, Josu},
  journal={Quantum},
  volume={8},
  pages={1498},
  year={2024},
  publisher={Verein zur F{\"o}rderung des Open Access Publizierens in den Quantenwissenschaften}
}

@article{berardini2025asymptotically,
  title={Asymptotically good {CSS-T} codes and a new construction of triorthogonal codes},
  author={Berardini, Elena and Dastbasteh, Reza and {Etxezarreta Martinez}, Josu and Jain, Shreyas and Larrarte, Olatz Sanz},
  journal={IEEE Journal on Selected Areas in Information Theory},
  year={2025},
  publisher={IEEE}
}

@book{huffman2010fundamentals,
  title={Fundamentals of error-correcting codes},
  author={Huffman, W Cary and Pless, Vera},
  year={2010},
  publisher={Cambridge university press}
}

@article{bravyi2012magic,
  title={Magic-state distillation with low overhead},
  author={Bravyi, Sergey and Haah, Jeongwan},
  journal={Physical Review A—Atomic, Molecular, and Optical Physics},
  volume={86},
  number={5},
  pages={052329},
  year={2012},
  publisher={APS}
}

@article{koutsioumpas2022smallest,
  title={The smallest code with transversal {T}},
  author={Koutsioumpas, Stergios and Banfield, Darren and Kay, Alastair},
  journal={arXiv preprint arXiv:2210.14066},
  year={2022}
}

@article{jones2013multilevel,
  title={Multilevel distillation of magic states for quantum computing},
  author={Jones, Cody},
  journal={Physical Review A—Atomic, Molecular, and Optical Physics},
  volume={87},
  number={4},
  pages={042305},
  year={2013},
  publisher={APS}
}

@article{heussen2025efficient,
  title={Efficient fault-tolerant code switching via one-way transversal {CNOT} gates},
  author={Heu{\ss}en, Sascha and Hilder, Janine},
  journal={Quantum},
  volume={9},
  pages={1846},
  year={2025},
  publisher={Verein zur F{\"o}rderung des Open Access Publizierens in den Quantenwissenschaften}
}

@article{jain2025transversal,
  title={Transversal {C}lifford and {T}-gate codes of short length and high distance},
  author={Jain, Shubham P and Albert, Victor V},
  journal={IEEE Journal on Selected Areas in Information Theory},
  year={2025},
  publisher={IEEE}
}

@inproceedings{leverrier2022quantum,
  title={{Quantum Tanner codes}},
  author={Leverrier, Anthony and Z{\'e}mor, Gilles},
  booktitle={2022 IEEE 63rd Annual Symposium on Foundations of Computer Science (FOCS)},
  pages={872--883},
  year={2022},
  organization={IEEE}
}

@article{haah2018codes,
  title={Codes and protocols for distilling {$T$}, controlled-{$S$}, and {T}offoli gates},
  author={Haah, Jeongwan and Hastings, Matthew B},
  journal={Quantum},
  volume={2},
  pages={71},
  year={2018},
  publisher={Verein zur F{\"o}rderung des Open Access Publizierens in den Quantenwissenschaften}
}

@article{hastings2018distillation,
  title={Distillation with sublogarithmic overhead},
  author={Hastings, Matthew B and Haah, Jeongwan},
  journal={Physical review letters},
  volume={120},
  number={5},
  pages={050504},
  year={2018},
  publisher={APS}
}

@article{webster2023transversal,
  title={Transversal diagonal logical operators for stabiliser codes},
  author={Webster, Mark A and Quintavalle, Armanda O and Bartlett, Stephen D},
  journal={New Journal of Physics},
  volume={25},
  number={10},
  pages={103018},
  year={2023},
  publisher={IOP Publishing}
}

@article{nezami2022classification,
  title={Classification of small triorthogonal codes},
  author={Nezami, Sepehr and Haah, Jeongwan},
  journal={Physical Review A},
  volume={106},
  number={1},
  pages={012437},
  year={2022},
  publisher={APS}
}

@article{leitch2025transversal,
  title={Transversal Gates for Highly Asymmetric {qLDPC} Codes},
  author={Leitch, Heather and Kay, Alastair},
  journal={arXiv preprint arXiv:2506.15905},
  year={2025}
}

@article{macwilliams1972good,
  title={Good self dual codes exist},
  author={MacWilliams, F Jessie and Sloane, Neil JA and Thompson, John G},
  journal={Discrete Mathematics},
  volume={3},
  number={1-3},
  pages={153--162},
  year={1972},
  publisher={Elsevier}
}

@article{rains2002self,
  title={Self-dual codes},
  author={Rains, Eric M and Sloane, Neil JA},
  journal={arXiv preprint math/0208001},
  year={2002}
}

@article{thompson1973weighted,
  title={Weighted averages associated to some codes},
  author={Thompson, JOHN G},
  journal={Scripta Math.},
  volume={29},
  pages={449},
  year={1973}
}

@article{camps2024algebraic,
  title={An algebraic characterization of binary {CSS-T} codes and cyclic {CSS-T} codes for quantum fault tolerance},
  author={Camps-Moreno, Eduardo and L{\'o}pez, Hiram H and Matthews, Gretchen L and Ruano, Diego and San-Jos{\'e}, Rodrigo and Soprunov, Ivan},
  journal={Quantum Information Processing},
  volume={23},
  number={6},
  pages={230},
  year={2024},
  publisher={Springer}
}

@article{randriambololona2015products,
  title={On products and powers of linear codes under componentwise multiplication},
  author={Randriambololona, Hugues},
  journal={Algorithmic arithmetic, geometry, and coding theory},
  volume={637},
  pages={3--78},
  year={2015},
  publisher={AMS Providence, RI, USA}
}

@article{dastbasteh2026quantum,
  title={Quantum Codes with Arbitrary {Z}-Rotation logical Gates and Applications to Fault-Tolerant Code Switching},
  author={Dastbasteh, Reza and Otxoa, Ruben M and Crespo, Pedro M and {Etxezarreta Martinez}, Josu},
  journal={arXiv preprint arXiv:2608.11160},
  year={2026}
}

@article{vuillot2022quantum,
  title={Quantum pin codes},
  author={Vuillot, Christophe and Breuckmann, Nikolas P},
  journal={IEEE Transactions on Information Theory},
  volume={68},
  number={9},
  pages={5955--5974},
  year={2022},
  publisher={IEEE}
}

@article{PhysRevA.92.022312,
  title = {Qudit color codes and gauge color codes in all spatial dimensions},
  author = {Watson, Fern H. E. and Campbell, Earl T. and Anwar, Hussain and Browne, Dan E.},
  journal = {Phys. Rev. A},
  volume = {92},
  issue = {2},
  pages = {022312},
  numpages = {15},
  year = {2015},
  month = {Aug},
  publisher = {American Physical Society},
  doi = {10.1103/PhysRevA.92.022312},
  url = {https://link.aps.org/doi/10.1103/PhysRevA.92.022312}
}

@article{krishna2018towards,
  title={Towards low overhead magic state distillation},
  author={Krishna, Anirudh and Tillich, Jean-Pierre},
  journal={arXiv preprint arXiv:1811.08461},
  year={2018}
}

@article{haah2018towers,
  title={Towers of generalized divisible quantum codes},
  author={Haah, Jeongwan},
  journal={Physical Review A},
  volume={97},
  number={4},
  pages={042327},
  year={2018},
  publisher={APS}
}

\appendix

\section{Distance two $r$-orthogonal codes} \label{A:generalfamily}
In Theorem \ref{T:4orthogonal} we introduced a family of 
$r$-orthogonal codes with parameters 
\[
[[2^{r}+(2^{r-1}-1)k,k,2]]
\]
for each even integer $k\ge 0$. 

The connection between the subclasses for $r=2,3$, namely $H$ and $[[3k+8,k,2]]$ families, was initially observed in \cite{jones2013multilevel}, and later it was shown that the triorthogonal $[[8+3k,k,2]]$ can be obtained from the H codes through a lifting process \cite{haah2018towers}. 
Here, we reveal a deeper connection through the language of simplex and Hamming codes, which also allows us to 
\begin{itemize}
    \item generalize the structure to any level of the Clifford hierarchy;
    \item show that they do not require a correction operator;
    \item calculate the error probabilities of the output magic states.
\end{itemize}

Our results are initially developed with more details for $r=4$, and then  generalized to larger $r$ values. 

Before presenting our results, we refer the reader to Section \ref{sec: hamming and simplex} for the notation and a quick overview of Hamming and simplex codes.

Let $H^\ast_3$ be the $3\times (2^{3}-1)$ parity-check matrix of the Hamming code of order $3$.  
Recall that $H^\ast_3$ is $2$-orthogonal. 
Consider the matrices
\[
S_1=\begin{bmatrix}
0&H^\ast_3\\
1&1
\end{bmatrix} \ \text{and} \  S_2=\begin{bmatrix}
H^\ast_3&H^\ast_3\\
0&0
\end{bmatrix}, 
\]
where $S_1$ is $4\times 2^3$ and $S_2$  is  $4\times 14$. 
Consider the block matrix 
\[
G=\begin{bmatrix}
S_1&S_1&S_2\\
0& L&M
\end{bmatrix},  
\]
where 
\[
M=\begin{bmatrix}
1&0\\
0&1
\end{bmatrix} \ \text{and} \
L=\begin{bmatrix}
1&1\\
1&1
\end{bmatrix}
\]
have sizes $2\times 14$ and $2\times 8$, respectively.

\begin{lemma}\label{L:fourOrtho}
The matrix $G$ is $4$-orthogonal, corresponding to a $[[30,2,2]]$ CSS code.    
\end{lemma} 

\begin{proof}
First, note that the last two rows of $G$ have weight $15$ and are orthogonal, corresponding to $[0 \ L \ M]$. Moreover, the rows of $G$ corresponding to $[S_1 \ S_1 \ S_2]$ all have even weights and all rows are $4$-orthogonal.  

Secondly, consider the Schur product of $4$ or fewer rows of $G$ consisting of the last two rows of $G$. Restricting to non-zero coordinates, the result of such a Schur product is the same as the Schur product of two rows of $S_1$. Finally, since $H^\ast_3$ is $2$-orthogonal, one can easily verify that $S_1$ is $2$-orthogonal as well. Hence, the result of the Schur product of such $4$ vectors always has even weight. 

The only remaining case is considering the Schur product of $4$ vectors (allowing for repetition) consisting of one row of $[0 \ L \ M ]$ and three rows of $[S_1 \ S_1 \ S_2]$. 
Again, restricting to non-zero coordinates results in the Schur product of $3$ or fewer vectors of 
\[
\begin{bmatrix}
    H^\ast_3 & H^\ast_3 &0\\
    0& 1 & 1
\end{bmatrix},
\]
which consists of all non-zero vectors of $\F_2^4$, that is $H^\ast_4$. This implies that the latter matrix is $3$-orthogonal and completes the argument about $4$-orthogonality of $G$.

The proof of the minimum distance is obvious. 
$\hfill \square$
\end{proof}

One can increase the number of logical qubits by adding more blocks as described below. 

\begin{theorem}\label{T:4orthogonal2}
There exists a family of $4$-orthogonal quantum CSS codes with parameters $[[16+7k,k,2]]$ for each even integer $k$.   
\end{theorem}

\begin{proof}
For each even $k$, one can form  a block matrix 
\[
G:=\begin{bmatrix}
S_1&S_1&S_2& S_2 &\cdots& S_2\\
0& L&M&0&\cdots& 0\\
0&L& 0&M&\cdots&0\\
\vdots &\vdots &\vdots &\vdots &\ddots&\vdots \\
0&L& 0&0&\cdots&M\\
\end{bmatrix},
\]
where there are $k/2$ blocks of $S_2$. The proof of $4$-orthogonality is by mimicking the idea of Lemma \ref{L:fourOrtho}, and we skip it here. 
The last $k$ rows of $G$ consist of odd-weight vectors, and the rest of the rows of $G$ have an even weight. The corresponding quantum code is defined as the CSS code of $C_2$ generated by the first block row of $G$ and $C_1$ generated by $G$.
One can easily confirm that any linear combination of the rows of $G$ has weight greater than or equal to two. Thus 
\[
d(C_1\setminus C_2)\ge 2.
\]
Moreover, there is no zero column in the matrix $G$, but the first row block of $G$ has two identical column blocks. This implies that 
\[
d(C_2^\bot \setminus C_1^\bot)\ge 2.
\]
Thus, such a CSS code has parameters 
\[
[[16+7k,k,2]].
\]
$\hfill \square$\end{proof}
Next, we generalize this result to $r$-orthogonality for each $r\ge 2$. 

\begin{theorem}\label{T:general family}
Let $r \ge 2$ be an integer and $k\ge 0$ be an even integer. 
Then there exists a family of quantum codes with parameters
\[
[[2^r+(2^{r-1}-1)k,k,2]]
\] 
that is $r$-orthogonal.

Moreover, such a quantum code admits a transversal logical 
$\bigotimes^k R_Z(\frac{-\pi}{2^{r-1}})
$ gate via the physical action
\[
\bigotimes^{2^{r-1}} R_Z\left(\frac{(-1)^\ell\pi}{2^{r-1}}\right) \bigotimes^{n-2^{r-1}} R_Z\left(\frac{\pi}{2^{r-1}}\right),
\]
where 
\[
\ell=\begin{cases}
0 & \ \text{if} \ k\equiv 2 \pmod 4 \\
1 & \ \text{if} \ k\equiv 0 \pmod 4. 
\end{cases}
\]
\end{theorem}

\begin{proof}
The proof is based on the process explained in the proof of Theorem \ref{T:4orthogonal2} and the discussion before it. 
So here we only provide a sketch for the proof.
We give the details of the logical action in Appendix \ref{A:gate part}.

Let $r\ge 2$ and $k\ge 0$ be fixed and $H_{r-1}^\ast$ be the parity check matrix of Hamming code of order $r-1$ (which has the size $(r-1)\times (2^{r-1}-1)$). 
As we mentioned earlier $H_{r-1}^\ast$ is $(r-2)$-orthogonal. 

Consider the matrix 
\begin{equation}\label{E:d2matrix}
G_r:=\begin{bmatrix}
S_1&S_1&S_2& S_2 &\cdots& S_2\\
0& L&M&0&\cdots& 0\\
0&L& 0&M&\cdots&0\\
\vdots &\vdots &\vdots &\vdots &\ddots&\vdots \\
0&L& 0&0&\cdots&M\\
\end{bmatrix},
\end{equation}
where there are $2$ blocks of $S_1$ and $k/2$ blocks of $S_2$ in the first row
\[
S_1=\begin{bmatrix}
0&H_{r-1}^\ast\\
1&1
\end{bmatrix} \ \text{and} \  S_2=\begin{bmatrix}
H_{r-1}^\ast&H_{r-1}^\ast\\
0&0
\end{bmatrix} 
\]
and 
\[
M=\begin{bmatrix}
1&0\\
0&1
\end{bmatrix} \ \text{and} \
L=\begin{bmatrix}
1&1\\
1&1
\end{bmatrix}
\]
with sizes $2\times 2(2^{r-1}-1)$ and $2\times 2^{r-1}$, respectively. 
The first block row of $G_r$ consists of even weight vectors ($X$-stabilizer basis), and the last $k$ block rows of $G_r$ consist of only odd weight vectors (basis of $X$-logical operators).
Because of the structure of $S_1$ and $S_2$, one can easily see that the first row block of $G_r$ is $r$-orthogonal.

Next, any Schur product of $r-1$ vectors from the first row block and one vector from the remaining row blocks is equivalent to the Schur product of $r-1$ vectors from 
\[
H_r^\ast=\begin{bmatrix}
 0&H_{r-1}^\ast&H_{r-1}^\ast\\ 
 1&1&0
\end{bmatrix}
\]
which is $(r-1)$-orthogonal. 
Next, if 2 or more different vectors outside the first row block appear in the Schur product, the product is equivalent to the Schur product of $r-2$
or less number of rows of $S_1$ (which is $(r-2)$-orthogonal). 
This proves the $r$-orthogonality of $G_r$. 

It is not difficult to see that the matrix $G_r$ has minimum weight $\ge 2$, and since the first row block ($X$-stabilizers) has two identical columns, the set of $Z$-normalizers has a weight-two vector. This shows that the minimum distance is two. 
Finally, there are $k$ rows with an odd weight corresponding to a basis of $k$ $X$-logical operators. 
This gives an $[[2^r+(2^{r-1}-1)k,k,2]]$.
$\hfill \square$
\end{proof}
  
Note also that the case $r=3$ is identical to the triorthogonal family given in \cite[Section VII]{bravyi2012magic}. The case $r=2$ also results in the self-orthogonal quantum codes known as the H codes, up to a permutation, where the connection is discussed in Appendix \ref{A:Hcode}.

The code families discussed in this section find applications in magic state distillation factories at level $r$ of the Clifford hierarchy. 
In the language of \cite{maleableTri}, the codes of Theorem \ref{T:general family} form distance-$2$ factories at level $r$ of the Clifford hierarchy with $k$ unentangled outputs. 
The two-group, degree-one factories found numerically in \cite[Table V]{maleableTri}, namely $[[30,2,2]]$, $[[44,4,2]]$, $[[58,6,2]]$ ($l=4$) match the parameters of codes produced via Theorem \ref{T:general family} for $k=2,4,6$. 
However, Theorem \ref{T:general family} gives a closed form family for every level and every even $k\ge 2$. 
Note also that for all codes with $k\equiv2 \pmod 4$ of Theorem \ref{T:general family} no correction would be required, though when $k \equiv 0 \pmod 4$ a simple sign adjustment, similar to those in \cite[Appendix F]{maleableTri}, can imply the desired transversal logical gate without a need for a correction operator. 
In Appendix \ref{A:gate part} below, we give more information about the structure of the logical operator, and also calculate the error probability in each output magic state.

\subsection{Realization of the transversal logical gate}\label{A:gate part}
In this section, we briefly discuss the choice of physical one-qubit operators for $r$-orthogonal codes of Theorem \ref{T:general family} to guarantee that the logical action is the transversal gate \[
\bigotimes^k R_Z\left(\frac{-\pi}{2^{r-1}}\right).
\]
The $r$-orthogonal matrix of an instance of such family with parameters $[[2^r+(2^{r-1}-1)k,k,2]]$ is represented in Equation \eqref{E:d2matrix}. 
Below we highlight some evident features of this matrix.
\begin{enumerate}
    \item The rows of the top blocks of  $S_1$ and $S_2$ have weights divisible by $2^{r-2}$ and $2^{r-1}$, respectively.
    \item The bottom block of $S_1$ is a vector with weight divisible by $2^{r-1}$.
    \item The Schur product of each $1\le s \le r-2$ number of distinct rows in $H_{r-1}^\ast$ has a weight divisible by $2^{r-1-s}$.
    \item All vectors representing logical $X$-operators in Equation \eqref{E:d2matrix} have weight $2^r-1$, and each two or more of the distinct such vectors overlap exactly at $2^{r-1}$ positions.  
    \item The Schur product of exactly one logical vector and one $X$-stabilizer can be expressed, after removing the zero blocks, as a vector in
    \[
H_r^\ast=\begin{bmatrix}
 0&H_{r-1}^\ast&H_{r-1}^\ast\\ 
 1&1&0
\end{bmatrix}.
\]
\item The Schur product of two or more distinct logical vectors and one $X$-stabilizer is equivalent to a vector of $S_1$.
\item When at least one logical vector appears in a Schur product, the resulting vector has all zeros in the first $S_1$ block.
\end{enumerate}
Now, we take advantage of the above properties and the result of Theorem \ref{T:logical con} to sketch how such a code implies a transversal logical 
$\bigotimes^k R_Z\left(\frac{-\pi}{2^{r-1}}\right)
$ gate via the physical action
\[
\bigotimes^{2^{r-1}} R_Z\left(\frac{(-1)^\ell\pi}{2^{r-1}}\right) \bigotimes^{n-2^{r-1}} R_Z\left(\frac{\pi}{2^{r-1}}\right),
\]
where all the negative signs impact the data qubits corresponding to the first $S_1$ block, and $\ell$ depends on $k$.

First, note that the conditions $(4)-(7)$ imply the desired logical action and the desired Schur product condition when at least one logical vector is involved. 
So we only concentrate on the $X$-stabilizer vectors.

The $X$-stabilizer matrix $H_x$ of such quantum code has 2 blocks of $S_1$ and $\frac{k}{2}$ blocks of $S_2$. 
Hence, the vectors belonging to the top block of $H_x$ consist of $k+2$ blocks of non-zero vectors, of the Simplex code $H_{r-1}^\ast$, all having weights divisible by $2^{r-2}$, and the bottom block is a vector with weight $2^r$.

We choose $\ell=0$ when $k\equiv 2 \pmod 4$, and $\ell=1$ when $k\equiv 0 \pmod 4$. Considering these signs, one can define a sign vector $t=(t_1,t_2,\ldots,t_n)$, where $t_i=(-1)^\ell$ for each $1\le i \le 2^{r-1}$ and the rest are $1$. 
Now the top and bottom blocks have vectors with weights divisible by $2^r$ (($\| \cdot\|_t \equiv 0 \pmod{2^r}$ in the sense of Theorem \ref{T:logical con}).  
Moreover, the Schur product of distinct vectors from blocks is equivalent to the Schur product of vectors of the form 
\[
\begin{bmatrix}
H_{r-1}^\ast&H_{r-1}^\ast
\end{bmatrix} 
\]
with both blocks having positive signs, or a positive and a negative sign. 
The structure of the above matrix implies the divisibility condition of Theorem \ref{T:logical con} for the Schur product of stabilizer vectors.

Therefore, the above lines assure that all the conditions of  Theorem \ref{T:logical con} are satisfied and the mentioned transversal gate is a valid logical operator.

\begin{remark}\label{rem:error probability}
Following the setup of \cite{bravyi2012magic}, one can assume that each of the single-qubit physical rotations is followed, independently with probability $p$, by a $Z$-type error on its qubit, and that all other operations are ideal.

All the weight two logical $Z$-operators are the $2^{r-1}$ pairs of equal columns in the two $S_1$ blocks and the $(2^{r-1}-1)\binom{k}{2}$ pairs of equal columns in the $S_2$ blocks. 
Hence, one can calculate the probability of ``at least one'' wrong output magic state as
$\big((2^{r-1}-1)\binom{k}{2}+2^{r-1}\big)p^2+\mathcal{O}(p^3)$. 
Moreover, every single output magic state has an error probability
$\big((2^{r-1}-1)k+1\big)p^2+\mathcal{O}(p^3)$, which reduces to the $(1+3k)p^2$ of \cite{bravyi2012magic} for $r=3$.   
\end{remark}

\section{H code as a subclass}\label{A:Hcode}
The family of $[[k+4,k,2]]$ H codes of \cite{jones2013multilevel} with even $k\ge 2$ is defined by the $X$-stabilizer operators 
\begin{equation}\label{E:1}    
X_1X_2X_3X_4 \ \text{and}\ X_1X_2X_5X_6\cdots X_{k+4}
\end{equation}
and $X$-logical operators 
\begin{equation}\label{E:2}
\overline{X_i}=X_1X_3X_{i+4}.  
\end{equation}
The set of $Z$-stabilizer and logical operators can be defined by exchanging $X$ with $Z$. 
Applying the convention of Remark \ref{R:Hadamard1}, we get $H_1^\ast=[1]$. 
Now applying the construction of Theorem \ref{T:general family} when $r=2$ implies the $X$-stabilizer operators
\[
X_2X_4X_5X_6\cdots X_{k+4} \ \text{and} \ X_1X_2X_3X_4
\]
and $X$-logical operators of the form 
\[
\overline{X_i}=X_3X_4X_{i+4}.  
\]
Now, swapping the operators in positions $1$ and $4$ of Equations \eqref{E:1} and \eqref{E:2} implies the above stabilizers and normalizers.

\section{Construction of matrices with two triorthogonal blocks}\label{A: two blocks}

Recall that we took advantage of matrices that are decomposable into two triorthogonal row blocks to build quantum triorthogonal codes with many logical qubits and potentially improve their minimum distance. 

A potential tool for forming a two-block triorthogonal decomposition was based on Proposition \ref{P:plotkin-two-triorthogonal}.
Next, we relax the condition of this result to construct a larger number of linear codes that are generated by two row block triorthogonal matrices. 

\begin{proposition}\label{P:plotkin-two-triorthogonal2}
Let $C_2=C_2^1\oplus C_2^2$ be a binary linear code such that both of $C_2^1$ and $C_2^2$ have even triorthogonal generator matrices. For any linear code $C_1$ such that $C_2^1\subseteq C_1$, we have that the code $P(C_1,C_2)$ has a generator matrix that consists of two triorthogonal row block matrices.  
\end{proposition}

\begin{proof}
Let $C_1=C_2^1\oplus C'$.
The code $P(C_1,C_2)$ has a generator matrix in the form 
\[
\begin{bmatrix}
C_2^1 & C_2^1\\
C' & C'\\
0 & C_2^1\\
0 & C_2^2\\
\end{bmatrix}.
\]
This generator matrix can be transformed to an equivalent one in the form 
\[
\begin{bmatrix}
C_2^1 & C_2^1\\
C' & C'\\
C_2^1 & 0\\
0 & C_2^2\\
\end{bmatrix}.
\]
Now choosing 
\[A=\begin{bmatrix}
C_2^1 & C_2^1\\
C' & C'\\
\end{bmatrix}\]
and 
\[
B=\begin{bmatrix}
C_2^1 & 0\\
0 & C_2^2\\
\end{bmatrix}
\] implies that $P(C_1,C_2)$ has the generator matrix $\begin{bmatrix}
    A\\B
\end{bmatrix}$ with $A$ and $B$ being both triorthogonal.\hfill$\square$
\end{proof}

\end{document}